\documentclass[usenatbib]{mn2e}

\newcommand\dosingle[1]{#1}  \newcommand\dodouble[1]{ } 

\newcommand\nice[1]{#1}    \newcommand\subm[1]{}   
\subm{ \renewcommand\dosingle[1]{ }   \renewcommand\dodouble[1]{#1} }

\dodouble{\documentclass[usenatbib,referee]{mn2e}}

\usepackage{graphicx}

\usepackage{amsfonts}

\usepackage{mn_bmath} 

\usepackage[varg]{txfonts}

\newcommand\mystamp[1]{#1}

\newcommand\mystamppreamble{
  \usepackage{eso-pic}
  \usepackage{color} 
  \definecolor{redstamp}{rgb}{0.99,0.80,0.90} 
  \usepackage{datetime}
  \usepackage[normalem]{ulem}
}

\mystamp{\mystamppreamble}

\usepackage{natbib}  
\usepackage{url}

\newcommand\prerefereechanges[1]{#1}

\newcommand\postrefereechanges[1]{#1}

\newcommand\postrefereechangesbis[1]{#1}

 \usepackage{hyperref}

\nice{ 
\providecommand{\eprint}[1]{\href{http://arxiv.org/abs/#1}{{\tt [arXiv:#1]}}}

\providecommand{\url}[1]{\href{#1}{#1}}

}

\providecommand{\adsurl}[1]{} 
\newcommand\SSS{Sect.~}

\providecommand\apj{ApJ}                 
\providecommand\aj{AJ}                 
\providecommand\apjs{ApJSupp}                 
\providecommand\apjl{ApJL}                 
\providecommand\aap{A\&A}            
\providecommand\mnras{MNRAS}

\providecommand\PRL{Physical Review Letters}
\providecommand\jetpL{Journal of Experimental and Theoretical Physics Letters}
\providecommand\PRL{PRL}
\providecommand\prd{PRD}
\providecommand\jetpL{JETPLett}
\providecommand\jrasc{J. Roy. Ast. Soc. Canada} 
\providecommand\jcap{JCAP}
\providecommand\ssr{Space~Sci.~Rev.}
\providecommand\physrep{Phys.Rep.}
\providecommand\pasj{PASJ}
\providecommand\nat{Nat.}
\providecommand\grg{Gen. Rel. Grav.}

\nice{ \newcommand\mycaptionfont{\protect\footnotesize} }

\newcommand\gtapprox{\,\lower.6ex\hbox{$\buildrel >\over \sim$} \, }
\newcommand\ltapprox{\,\lower.6ex\hbox{$\buildrel <\over \sim$} \, }
\newcommand\propapprox{\,\lower.6ex\hbox{$\buildrel \propto\over \sim$} \, }

\newcommand\arcs{\ifmmode {'' }\else $'' $\fi}     
\newcommand\arcm{\ifmmode {' }\else $' $\fi}       
\newcommand\ddeg{\ifmmode^\circ\else$^\circ$\fi}    

\newcommand\diffd{\mathrm{d}}

\newcommand\frtoday{Le\space\number\day\space\ifcase\month\or
  janvier\or f\'evrier\or mars\or avril\or mai\or juin\or
  juillet\or ao\^ut\or septembre\or octobre\or novembre\or 
d\'ecembre\fi\space \number\year}
\def\frdutoday{du\space\number\day\space\ifcase\month\or
  janvier\or f\'evrier\or mars\or avril\or mai\or juin\or
  juillet\or ao\^ut\or septembre\or octobre\or novembre\or 
d\'ecembre\fi\space \number\year}

\newcommand\todayISO{\number\year-\ifnum\month<10 0\fi\number\month-\ifnum\day<10 0\fi\number\day}

\newcommand\cqg{ClassQuantGra}   %
\newcommand\BASI{Bull. Astr. Soc. India}

\newcommand\hGpc{\mbox{$h^{-1}$ Gpc}}
\newcommand\hMpc{\mbox{$h^{-1}$ Mpc}}
\newcommand\hMpccube{\mbox{$h^3$ Mpc$^{-3}$}}

\newcommand\hGyr{\mbox{$h^{-1}$ Gyr}}
\newcommand\hMyr{\mbox{$h^{-1}$ Myr}}

\newcommand\rSLS{r_{\mathrm{SLS}}}  

\newcommand\Omm{\Omega_{\mathrm{m}}}
\newcommand\Omrad{\Omega_{\mathrm{r}}}

\newcommand\ximc{\xi_{\mathrm{C}}} 

\newcommand\zcosm{{z}_{\mathrm{cosm}}} 
\newcommand\zpec{{z}_{\mathrm{pec}}} 
\newcommand\phicosm{{\phi}_{\mathrm{cosm}}} 
\newcommand\phipec{{\phi}_{\mathrm{pec}}} 
\newcommand\betacosm{{\beta}_{\mathrm{cosm}}} 
\newcommand\betapec{{\beta}_{\mathrm{pec}}} 
\newcommand\zphot{z_{\mathrm{phot}}} 
\newcommand\gammacosm{{\gamma}_{\mathrm{cosm}}} 
\newcommand\gammapec{{\gamma}_{\mathrm{pec}}} 
\newcommand\epsBoP{\epsilon_{\mathrm{BoP}}}

\newcommand\lTthr{l^{T^3}}
\newcommand\bTthr{b^{T^3}}
\newcommand\LTthr{L^{T^3}}

\newcommand\probbeta{p_{0}}
\newcommand\probbetabest{p_{1}}

\providecommand\tablefoot[1]{{#1}}
\providecommand\tablefoottext[1]{${}^{#1}$}
\providecommand\tablefootmark[1]{${}^{#1}$\,\ignorespaces}
\providecommand\tablefootmarkmath[1]{^{#1}\,\ignorespaces}

\begin{document}

\title[Deep redshift topological lensing: $T^3$]{Deep redshift topological lensing: strategies for the $T^3$ candidate}

\author[Roukema et al.]{Boudewijn F. Roukema$^{1,2}$,
  Martin J. France$^{2}$,
  Tomasz A. Kazimierczak$^{1}$ \&
  Thomas Buchert$^{2}$
  \\
  $^1$ Toru\'n Centre for Astronomy, 
  Faculty of Physics, Astronomy and Informatics,
  Nicolaus Copernicus University,
  ul. Gagarina 11, 87-100 Toru\'n, Poland 
  \\
  $^2$
  Universit\'e de Lyon, Observatoire de Lyon,
  Centre de Recherche Astrophysique de Lyon, CNRS UMR 5574: Universit\'e Lyon~1 and \'Ecole Normale\\ Sup\'erieure de Lyon, 
  9 avenue Charles Andr\'e, F--69230 Saint--Genis--Laval, France\protect\thanks{BFR: during visiting lectureship.}}


\date{\frtoday}



\newcommand\Nchainsmain{16}
\newcommand\Npergroup{four}

\maketitle

\begin{abstract}
{The 3-torus ($T^3$) Friedmann-Lema\^{\i}tre-Robertson-Walker (FLRW)
  model better fits the nearly zero large-scale two-point
  auto-correlation of the Wilkinson Microwave Anisotropy Probe (WMAP)
  cosmic microwave background (CMB) sky maps than the infinite flat
  model.  The $T^3$ model's parameters, recently found using an
  optimal cross-correlation method on WMAP data, imply approximately
  equal-redshift topological lensing at redshifts $z \sim 6$, the
  redshift range of the upcoming generation of new instruments and
  telescopes.}
{We investigate observational strategies that can reject the $T^3$
  solution for a given region of parameter space of physical
  assumptions, or provide good candidate topologically lensed galaxy
  pairs for detailed spectroscopic followup.}
{$T^3$ holonomies are applied to (i) existing $z\sim 6$ observations and
  (ii) simulated observations, creating multiply connected catalogues.
  Corresponding simply connected catalogues are generated. 
  The simulated observational strategies are motivated by 
  the matched discs principle. Each catalogue is analysed
  using a successive filter method and collecting matched quadruples.
  Quadruple statistics between the multiply and simply connected
  catalogues are compared.
}
{The expected rejection of the hypothesis, or detection of candidate
  topologically lensed galaxies, is possible at a significance of 5\% 
  for a pair of $T^3$ axis-centred northern and southern surveys 
  if photometric redshift
  accuracy is $\sigma(\zphot) \ltapprox 0.01$ for a pair of 
  nearly complete 100~deg$^2$ 
  surveys with a total of $\gtapprox 500$ galaxies over $4.3 < z < 6.6$, or
  for a pair of 196~deg$^2$  surveys with $\gtapprox 400$ galaxies 
  and $\sigma(\zphot) \ltapprox 0.02$ over $4<z<7$. Dropping the maximum 
  time interval in a pair from $\Delta t =1$~{\hGyr} to 
  $\Delta t =0.1$~{\hGyr} requires
  $\sigma(\zphot) \ltapprox 0.005$ or
  $\sigma(\zphot) \ltapprox 0.01$, respectively.
}
{Millions of $z \sim 6$ galaxies will be observed over fields of these
  sizes during the coming decades, implying much stronger
  constraints. The question is not {{\em if}} the hypothesis will be
  rejected or confirmed, it is {{\em when}}.}
\end{abstract}


\mystamp{}


\dodouble{ \clearpage } 


\newcommand\tsubaru{
\begin{table*}
\caption{\mycaptionfont Known objects\tablefootmark{a} near the
  northern galactic matched disc at sky position $(l_1,b_1)$ and redshift $z_1$
  and their predicted most likely southern galactic positions $(l_2,b_2)$ and redshifts
  $z_2$, 
  and redshift and cosmological time differences 
  $\Delta z := z_1 - z_2$ and
  $\Delta t := t_2 - t_1$ (in {\hMyr}), respectively,
  assuming metric parameters $\Omm=0.28, \Omrad = 1.65 \times
  10^{-4} \Omm, \Omega_\Lambda = 1 - \Omm - \Omrad$ and the
  $T^3$ fundamental axis $(l=6^\circ, b=+77^\circ)$ of length
  $L=11.4{\hGpc}$.  
\label{t-subaru}}
$$\begin{array}{l ccc ccc rr} \hline \hline
\rule{0ex}{2.5ex}
\mbox{name} 
& 
l_1 & b_1 & 
l_2 & b_2 & 
z_1 & z_2 & \Delta z & \Delta t
\\  \hline
\rule{0ex}{2.5ex}
\mbox{{[SKD2006]~017721}} &     35.597 &     82.506 &    174.850 &    -70.004 &      5.724 &      5.281 &      0.443 &       75.6 \\ 
\mbox{{[SKD2006]~036334}} &     36.053 &     82.455 &    174.632 &    -70.020 &      5.705 &      5.300 &      0.405 &       69.2 \\ 
\mbox{{[SKD2006]~034503}} &     36.000 &     82.515 &    174.771 &    -69.984 &      5.698 &      5.308 &      0.390 &       66.6 \\ 
\mbox{SDF~J132519.4+271829} &     36.238 &     82.470 &    174.368 &    -69.758 &      6.000 &      5.058 &      0.942 &      160.2 \\ 
\mbox{[SKD2006]~018699} &     35.605 &     82.619 &    175.130 &    -69.980 &      5.645 &      5.356 &      0.289 &       49.4 \\ 
\mbox{[SKD2006]~020087} &     35.642 &     82.613 &    175.045 &    -69.917 &      5.724 &      5.286 &      0.438 &       74.8 \\ 
\mbox{[SKD2006]~031858} &     35.934 &     82.554 &    174.870 &    -69.965 &      5.692 &      5.314 &      0.378 &       64.5 \\ 
\mbox{SDSS~J134040.24+281328.1}\tablefootmarkmath{b} &     41.523 &     79.054 &    165.545 &    -71.950 &      5.338 &      5.654 &     -0.315 &      -53.9 \\ 
\mbox{[SKD2006]~020495} &     35.635 &     82.719 &    175.272 &    -69.858 &      5.697 &      5.314 &      0.383 &       65.4 \\ 
\mbox{SDF~J132426.5+271600} &     35.900 &     82.666 &    174.836 &    -69.609 &      6.030 &      5.039 &      0.991 &      168.2 \\ 
\mbox{[SKD2006]~023759} &     35.721 &     82.737 &    175.288 &    -69.846 &      5.687 &      5.324 &      0.363 &       61.9 \\ 
\mbox{[SKD2006]~031765} &     35.920 &     82.705 &    175.167 &    -69.852 &      5.691 &      5.321 &      0.370 &       63.1 \\ 
\mbox{[SKD2006]~073078} &     37.030 &     82.493 &    174.428 &    -69.913 &      5.700 &      5.314 &      0.386 &       65.8 \\ 
\mbox{[SKD2006]~027787} &     35.814 &     82.765 &    175.300 &    -69.801 &      5.707 &      5.308 &      0.399 &       68.0 \\ 
\mbox{[SKD2006]~046904} &     36.307 &     82.703 &    175.075 &    -69.843 &      5.665 &      5.347 &      0.318 &       54.2 \\ 
\mbox{[SKD2006]~042576} &     36.182 &     82.745 &    175.175 &    -69.806 &      5.684 &      5.331 &      0.353 &       60.2 \\ 
\mbox{[SKD2006]~039849} &     36.112 &     82.767 &    175.232 &    -69.788 &      5.692 &      5.324 &      0.368 &       62.7 \\ 
\mbox{[SKD2006]~050215} &     36.394 &     82.720 &    175.070 &    -69.809 &      5.682 &      5.333 &      0.349 &       59.4 \\ 
\mbox{SDF~J132501.3+272628} &     37.258 &     82.538 &    174.767 &    -70.166 &      5.337 &      5.666 &     -0.329 &      -56.1 \\ 
\mbox{SDF~J132442.5+272423} &     36.992 &     82.607 &    174.394 &    -69.554 &      6.040 &      5.038 &      1.002 &      169.8 \\ 
\mbox{SDF~J132345.6+271701} &     36.019 &     82.817 &    175.043 &    -69.415 &      6.110 &      4.986 &      1.124 &      190.4 \\ 
\mbox{[SKD2006]~061418} &     36.716 &     82.675 &    174.839 &    -69.763 &      5.748 &      5.276 &      0.472 &       80.3 \\ 
\mbox{[SKD2006]~048328} &     36.338 &     82.782 &    175.196 &    -69.758 &      5.693 &      5.325 &      0.368 &       62.6 \\ 
\mbox{[SKD2006]~093966} &     37.603 &     82.523 &    174.308 &    -69.824 &      5.721 &      5.302 &      0.419 &       71.4 \\ 
\mbox{[SKD2006]~070600} &     36.978 &     82.784 &    175.054 &    -69.738 &      5.654 &      5.366 &      0.288 &       49.0 \\ 
\mbox{[SKD2006]~096007} &     37.679 &     82.641 &    174.544 &    -69.752 &      5.696 &      5.329 &      0.367 &       62.5 \\ 
\mbox{[SKD2006]~124783} &     38.436 &     82.458 &    173.942 &    -69.801 &      5.720 &      5.308 &      0.412 &       70.1 \\ 
\mbox{[SKD2006]~095588} &     37.672 &     82.650 &    174.530 &    -69.710 &      5.738 &      5.292 &      0.446 &       75.8 \\ 
\mbox{[SKD2006]~084305} &     37.352 &     82.740 &    174.832 &    -69.706 &      5.695 &      5.331 &      0.364 &       62.0 \\ 
\mbox{[SKD2006]~124530} &     38.442 &     82.515 &    174.085 &    -69.784 &      5.689 &      5.337 &      0.352 &       59.8 \\ 
\mbox{[SKD2006]~084720} &     37.374 &     82.768 &    174.891 &    -69.695 &      5.682 &      5.343 &      0.339 &       57.6 \\ 
\mbox{[SKD2006]~132343} &     38.639 &     82.481 &    173.954 &    -69.785 &      5.697 &      5.331 &      0.366 &       62.3 \\ 
\mbox{[SKD2006]~121315} &     38.376 &     82.624 &    174.302 &    -69.688 &      5.716 &      5.317 &      0.399 &       67.8 \\ 
\mbox{[SKD2006]~113271} &     38.176 &     82.675 &    174.464 &    -69.673 &      5.710 &      5.322 &      0.388 &       65.9 \\ 
\mbox{[SKD2006]~126848} &     38.522 &     82.593 &    174.197 &    -69.697 &      5.718 &      5.315 &      0.403 &       68.4 \\ 
\mbox{[SKD2006]~108164} &     38.033 &     82.762 &    174.709 &    -69.653 &      5.672 &      5.358 &      0.314 &       53.4 \\ 
\mbox{[SKD2006]~152586} &     39.251 &     82.713 &    174.280 &    -69.578 &      5.681 &      5.359 &      0.322 &       54.7 \\ 
\mbox{[SKD2006]~154296} &     39.280 &     82.709 &    174.235 &    -69.550 &      5.715 &      5.329 &      0.386 &       65.5 \\ 
\mbox{[SKD2006]~166310} &     39.595 &     82.696 &    174.140 &    -69.547 &      5.698 &      5.346 &      0.352 &       59.6 \\ 
\mbox{[SKD2006]~158036} &     39.404 &     82.755 &    174.329 &    -69.540 &      5.676 &      5.366 &      0.310 &       52.5 \\ 
\mbox{[SKD2006]~168127} &     39.668 &     82.757 &    174.268 &    -69.518 &      5.674 &      5.370 &      0.304 &       51.4 \\ 
\mbox{[SKD2006]~163079} &     39.556 &     82.785 &    174.353 &    -69.505 &      5.676 &      5.368 &      0.308 &       52.1 \\ 
\mbox{SDSS~J133412.56+122020.7}\tablefootmarkmath{b} &    339.134 &     72.136 &    223.205 &    -77.385 &      5.134 &      5.962 &     -0.828 &     -139.6 \\ 
\mbox{SDSS~J140940.72+274507.1}\tablefootmarkmath{c} &     39.513 &     72.652 &    146.699 &    -75.339 &      5.359 &      5.819 &     -0.460 &      -76.1 \\ 
\hline
\end{array} $$
\tablefoot{ 
  \tablefoottext{a}{The objects are galaxies except where
    otherwise indicated.}  
  \tablefoottext{b}{QSO.}
  \tablefoottext{c}{Object type uncertain.}
}
\end{table*}
}  

\newcommand\tTthrlb{
\begin{table}
\caption{\mycaptionfont Preferred fundamental directions
  $1\le i \le 6$
  of the \protect\citet{Aurich08a} $T^3$ solution 
  in galactic coordinates\tablefootmark{a},
  with a directional uncertainty of $\sim 2^\circ$ (great circle degrees).
  \label{t-T3-lb}}
      $$\begin{array}{c c rrr rrr} \hline\hline
        & i  & 1 & 2 & 3 & 
        4\tablefootmarkmath{b} & 
        5\tablefootmarkmath{b} & 
        6\tablefootmarkmath{b} \\ 
        &\mathrm{unit} \\
        \hline 
        \lTthr_i & ^{\circ} & 6  &  17 & 107 &  186 & 197 & 287 \\
        \bTthr_i & ^{\circ} & 77 & -13 &   3 &  -77 &  13 &  -3 \\
        \LTthr & c/H_0 & \multicolumn{6}{c}{3.80 \pm 0.05\tablefootmarkmath{c}} \\
        \hline
        \end{array}$$ \\
\tablefoot{ 
  \tablefoottext{a}{The typographical error (sign of
    $\bTthr_6$) in Table~1 of \protect\citet{Aurich08a} has been
    corrected (cf Figs~6,~7 of \protect\citealt{Aurich08a}).}\\
  \tablefoottext{b}{The directions 4, 5, and 6 are exactly antipodal to the directions 1, 2, and 3, 
    respectively, and are listed for convenience only.}\\
  \tablefoottext{c}{Mean and standard error adopted from our analysis (\protect\ref{s-method-T3-WMAP}).}
}
\end{table}
}  

\newcommand\tzmin{
\begin{table}
\caption{\mycaptionfont Redshifts increasing radially
  outwards from the centre of matched discs for the 
 \protect\citet{Aurich08a} $T^3$ solution, but with
 $\LTthr=3.8$ motivated by the WMAP7 ILC map
 (Fig.~\protect\ref{f-xi-ilc-85}), at angle
 $\beta$ from the disc centre, and approximate fraction of full sky
 covered by the three pairs of discs $\omega/(4\pi)$\tablefootmark{a},
 depending on the assumed FLRW metric parameter $\Omm$, where 
 $\Omega_\Lambda = 1 -\Omm$.
  \label{t-zmin}}
      $$\begin{array}{r rrrrr r} 
  \hline \hline \rule{0ex}{2.5ex} 
          \Omm =  & 0.26 &0.27 &  0.28 &  0.29 & 0.30 \\
          \beta & \multicolumn{5}{c}{z} &\omega/(4\pi) \\
          \hline\rule{0ex}{2.5ex} 
          {}^\circ \\
          \hline\rule{0ex}{2.5ex} 
0 & 4.98 & 5.17 & 5.35 & 5.55 & 5.75 &
           0\% \\
5 & 5.04 & 5.22 & 5.42 & 5.61 & 5.82 &
           1\% \\
10 & 5.21 & 5.41 & 5.61 & 5.82 & 6.04 &
          5\% \\
\mbox{\em 10}\tablefootmarkmath{b} & 
\mbox{\em 6.10} & 
\mbox{\em 6.36} & 
\mbox{\em 6.63} & 
\mbox{\em 6.92} & 
\mbox{\em 7.21} &
\mbox{\em   5\%} \\
15 & 5.52 & 5.74 & 5.96 & 6.20 & 6.45 &
          10\% \\
20 & 6.02 & 6.27 & 6.53 & 6.81 & 7.10 &
          18\% \\
25 & 6.78 & 7.1 & 7.42 & 7.77 & 8.13 &
          28\% \\
30 & 7.99 & 8.41 & 8.85 & 9.31 & 9.81 &
          40\% \\
          \hline
      \end{array}$$
\tablefoot{ 
  \tablefoottext{a}{Galactic masking would modify these rough estimates.} \\
  \tablefoottext{b}{Calculation assuming $\LTthr=4.0$ for the circle with $\beta=10^\circ$, to illustrate the effect of uncertainty in $\LTthr$.} 
}
\end{table}
}  

\newcommand\tTthrsofar{
\begin{table}
\caption{\mycaptionfont Number of astrophysical objects with $5 \le z
  \le 6.2$ and within 10$^\circ$ of the \protect\citet{Aurich08a}
  $T^3$ solution fundamental axes 
  (numbered as in Table~\protect\ref{t-T3-lb}) 
  listed in NED\tablefootmark{a} as of 28 Oct 2012.
  \label{t-T3-so-far}}
$$\begin{array}{c  rrr rrr} \hline\hline
  i & 1 & 2 & 3 &   4 & 5 & 6 \\
  \hline 
  \mbox{NED count} & 44 & 0 & 0 &   9 & 1 & 0 \\
  \hline
\end{array}$$ \\
\tablefoot{
  \tablefoottext{a}{NASA/IPAC Extragalactic Database,
    (\protect\url{http://ned.ipac.caltech.edu})}
}
\end{table}
}  

\newcommand\tdeepfields{
\begin{table}
\caption{\mycaptionfont Some well-known deep fields and their
  angular distance $\theta_i$ from the nearest $T^3$ axis (listed if 
  $\theta \le 20^\circ$).
  \label{t-deepfields}}
      $$\begin{array}{r rrrrr} 
  \hline \hline \rule[-1.1ex]{0ex}{3.2ex}
        \mathrm{name} & l
        & b
        & \mathrm{min}_{i=1}^6(\theta_i) 
        & l_{T^3} & b_{T^3}\\
        \mathrm{unit} & {}^\circ         & {}^\circ         & {}^\circ         & {}^\circ         & {}^\circ \\
        \hline 
        \rule{0ex}{2.5ex}
        \mathrm{VVDS}~\mbox{0226$-$04}\tablefootmarkmath{a} &              172.&   -58.&    20.&   186.&    -77. \\
        \mathrm{CDFS}\tablefootmarkmath{b} &                               224.&   -54.&    26.\\
        \mbox{HDF-S}\tablefootmarkmath{c} &                                328.&   -49.&    52.\\
        \mathrm{VVDS}~\mbox{2217+00}\tablefootmarkmath{a} &                 63.&   -44.&    50.\\
        \mathrm{VVDS}~\mbox{1003+01}\tablefootmarkmath{a} &                238.&    43.&    46.\\
        \mathrm{HDF}\tablefootmarkmath{d} &                                126.&    55.&    43.\\
        \mathrm{AEGIS}\tablefootmarkmath{e} &                               97.&    60.&    33.\\
        \mathrm{VVDS}~\mbox{1400+05}\tablefootmarkmath{f} &                343.&    63.&    16.&     6.&     77. \\
        \mathrm{SDF}\tablefootmarkmath{g} &                                 38.&    83.&     8.&     6.&     77. \\
        \hline
      \end{array}$$
\tablefoot{ 
  \tablefoottext{a}{VIRMOS Very Deep Survey \protect\citep{VVDS03MNRAS}} \\
  \tablefoottext{b}{\mbox{Chandra Deep Field South} 
    (\protect\postrefereechangesbis{\protect\citealt{GiacconiCFDS01}};\\ 
    {\protect\url{http://www.eso.org/~vmainier/cdfs_pub/CDFSfield.html}})}\\
  \tablefoottext{c}{Hubble Deep Field-South ``WPFC2 (apex) pointing'' 
    (\protect\postrefereechangesbis{\protect\citealt{HDFS00MNRAS}};\\ 
    {\protect\url{http://www.stsci.edu/ftp/science/hdfsouth/coordinatesS.html}})} \\
  \tablefoottext{d}{Hubble Deep Field \protect\citep{HDFN96}}\\
  \tablefoottext{e}{\protect\postrefereechangesbis{All-Wavelength Extended Groth Strip International Survey \protect\citep{AEGIS07MNRAS}}}\\
  \tablefoottext{f}{\protect\citet{VVDSIovino05MNRAS}} \\
  \tablefoottext{g}{Subaru Deep Field \protect\citep{SubaruDF01MNRAS}}
}
\end{table}
}  

\newcommand\fxiilcmain{
\begin{figure}
\centering 
\includegraphics[width=8cm]{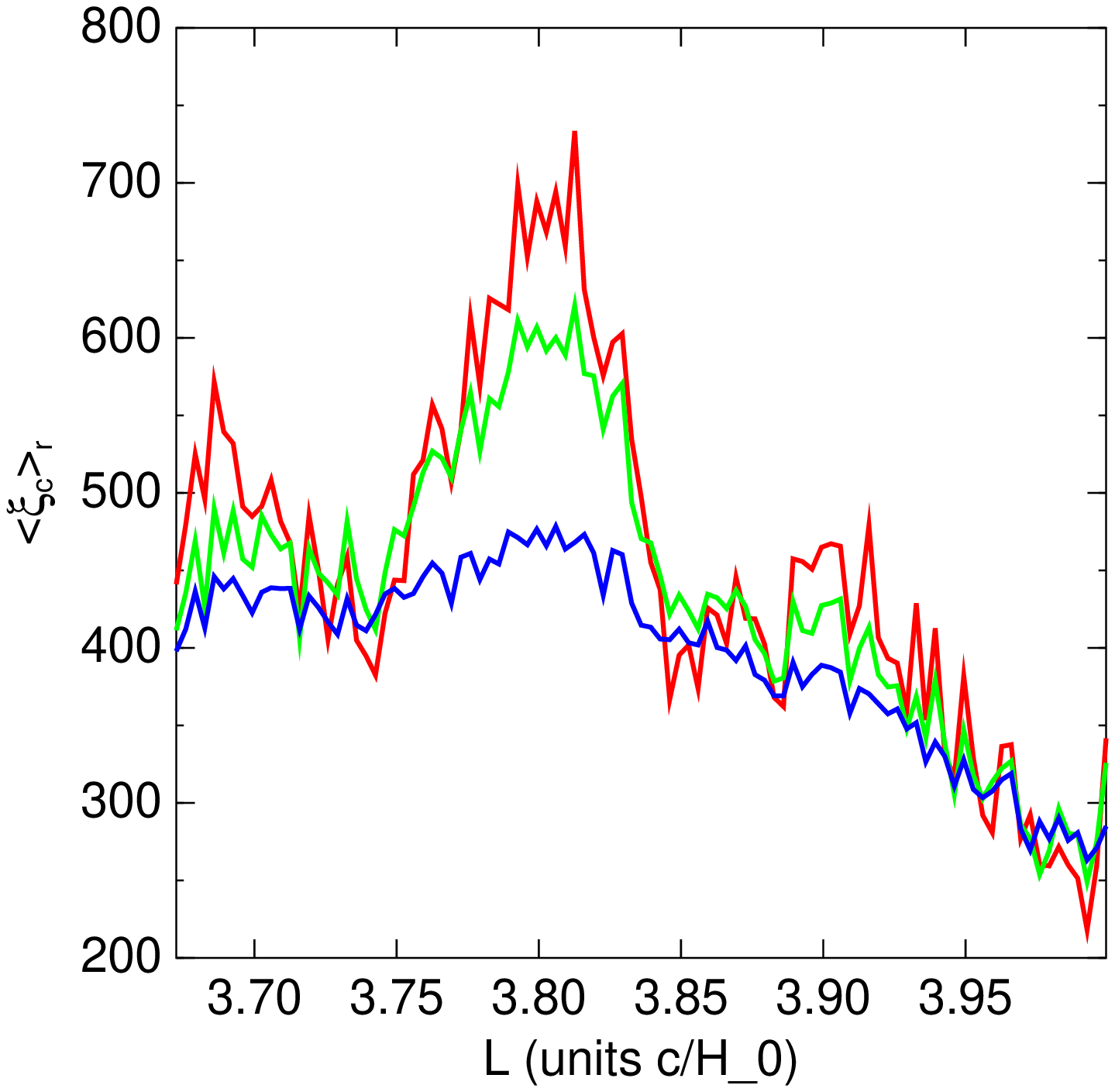}
\caption[]{ \mycaptionfont 
Sub-gigaparsec-scale mean cross-correlation 
\prerefereechanges{$\left<\ximc\right>_r(L)$} [Eq.~(\protect\ref{e-defn-xim})] 
{in $\mu \mathrm{K}^2$}
as a function of Universe size $\LTthr$ in units of $c/H_0$,
for the 
WMAP 9-year 
ILC map
using the KQ85 galactic mask, for $r=0.2, 0.4, 1.0\,{\hGpc}$
from top to bottom (red, green, blue, respectively, online),
for the orientation given in Table~\protect\ref{t-T3-lb}.
}
\label{f-xi-ilc-85}
\end{figure} 
} 

\newcommand\fmatchedbeams{
\begin{figure}
\centering 
\includegraphics[width=1.0\columnwidth]{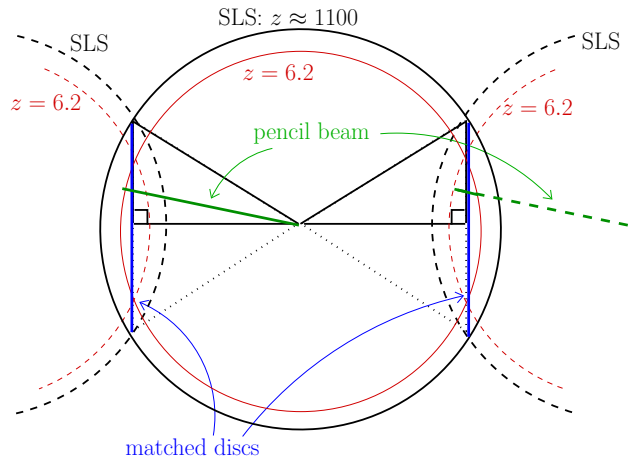} 
\caption[]{ \mycaptionfont {Matched discs in the universal
covering space, as per Fig.~1, \protect\citet{RK11},
but for the \protect\citet{Aurich08a} $T^3$ solution, with
the addition of matched pencil beam observations. Spheres 
corresponding to an example redshift $z=6.2$ are shown intersecting
themselves (in the covering space). 
The redshift at the centres of the matched discs is 
less than this, and within a disc, the redshift increases radially out
to the surface of last scattering (SLS). An observer (at the 
centre) pointing a telescope towards the right-hand matched disc
can observe the $z \ltapprox 6.2$ portion of the right-hand
copy of the pencil beam at roughly the same redshifts as those
of the corresponding portion of the left-hand copy of the pencil beam.
\label{f-matched-beams}
}}
\end{figure} 
}

\newcommand\fquadruples{
  \begin{figure}
    \centering 
    \includegraphics[width=0.8\columnwidth]{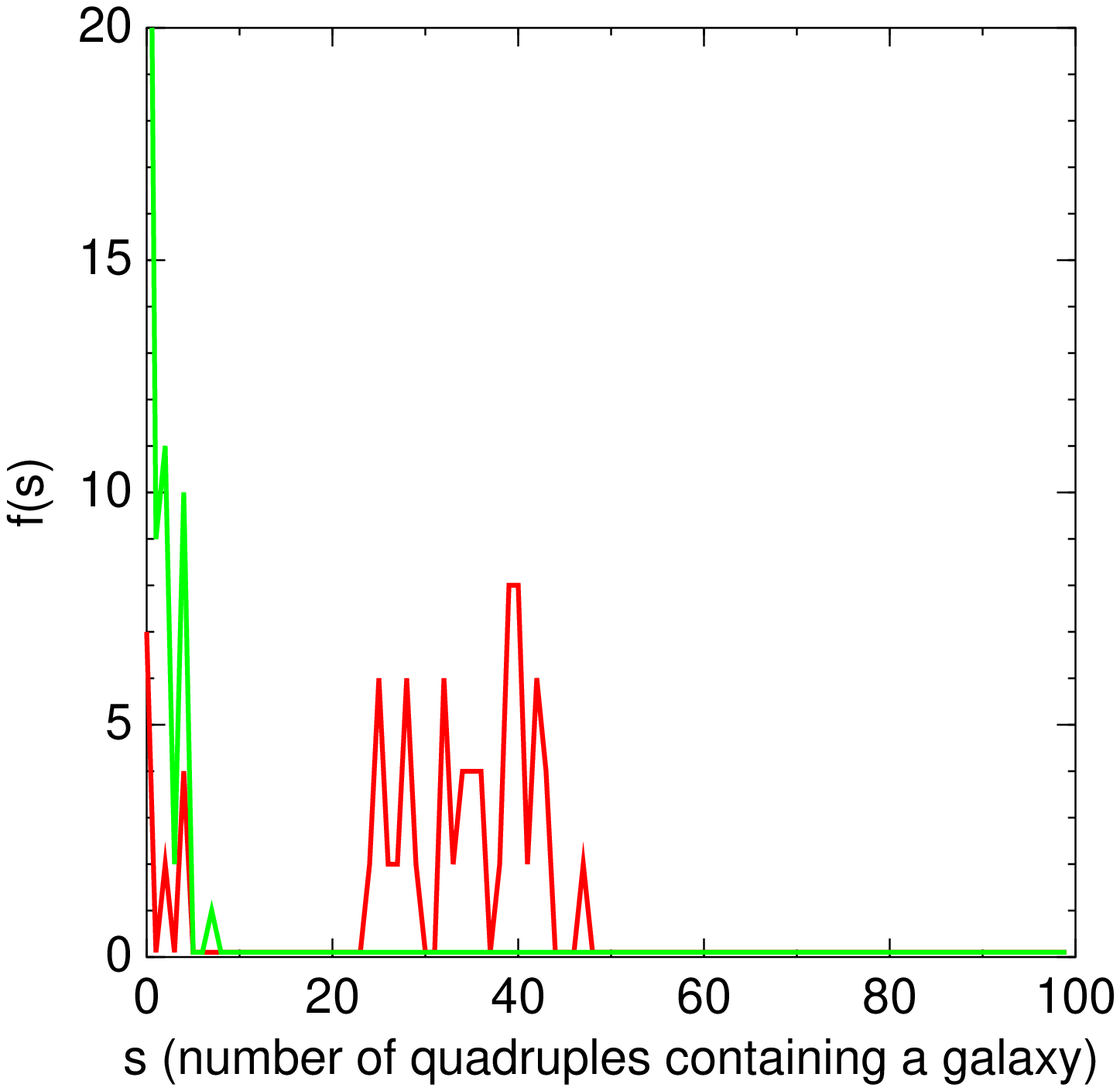} 
    \includegraphics[width=0.8\columnwidth]{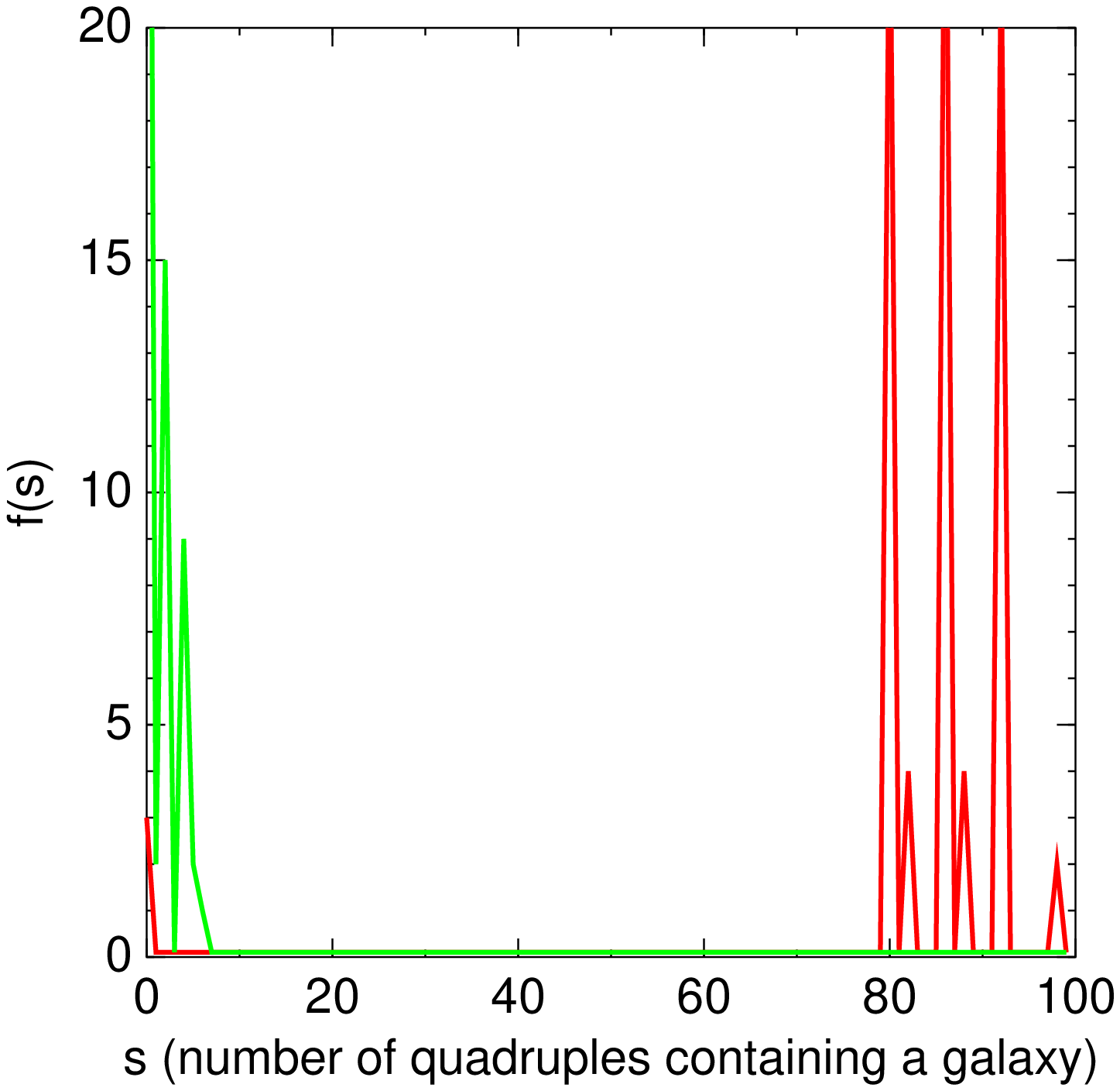} 
    \caption[]{ \mycaptionfont {Quadruple frequencies $N(s)$ obtained
        by the successive filter method
        (\SSS\protect\ref{s-method-filters}) applied to sets of
        objects in the multiply connected (dark curve, red online) and
        simply connected (pale curve, green online) cases, using the
        known, northern galactic objects in the SDF (second, third and
        sixth columns of Table~\protect\ref{t-subaru}). The southern
        object set is calculated \protect\prerefereechanges{using} 
        the \protect\citet{Aurich08a} $T^3$
        model in the multiply connected case, and simulated with an
        input auto-correlation function $r_0 = 10.0{\hMpc}, \gamma =
        1.8, r_{\min} = 1.0{\hMpc}, r_{\max} = 100.0{\hMpc}$ in the
        simply connected case.  The southern field is ``observed''
        over 1~deg$^2$ at the expected position in both cases.  {\em
          Top:} Successive filter parameters
        (\SSS\protect\ref{s-method-filters}) are $\epsilon = 1.0
        {\hMpc}, \Delta t = 0.01 {\hGyr}, \epsBoP = 1.0 {\hMpc}$.
        {\em Bottom:} Same, except that $\Delta t = 1 {\hGyr}$.
        \label{f-quadruples}
    }}
  \end{figure} 
} 

\newcommand\fquadruplesTthreeperturb{
  \begin{figure}
    \centering 
    \includegraphics[width=0.8\columnwidth]{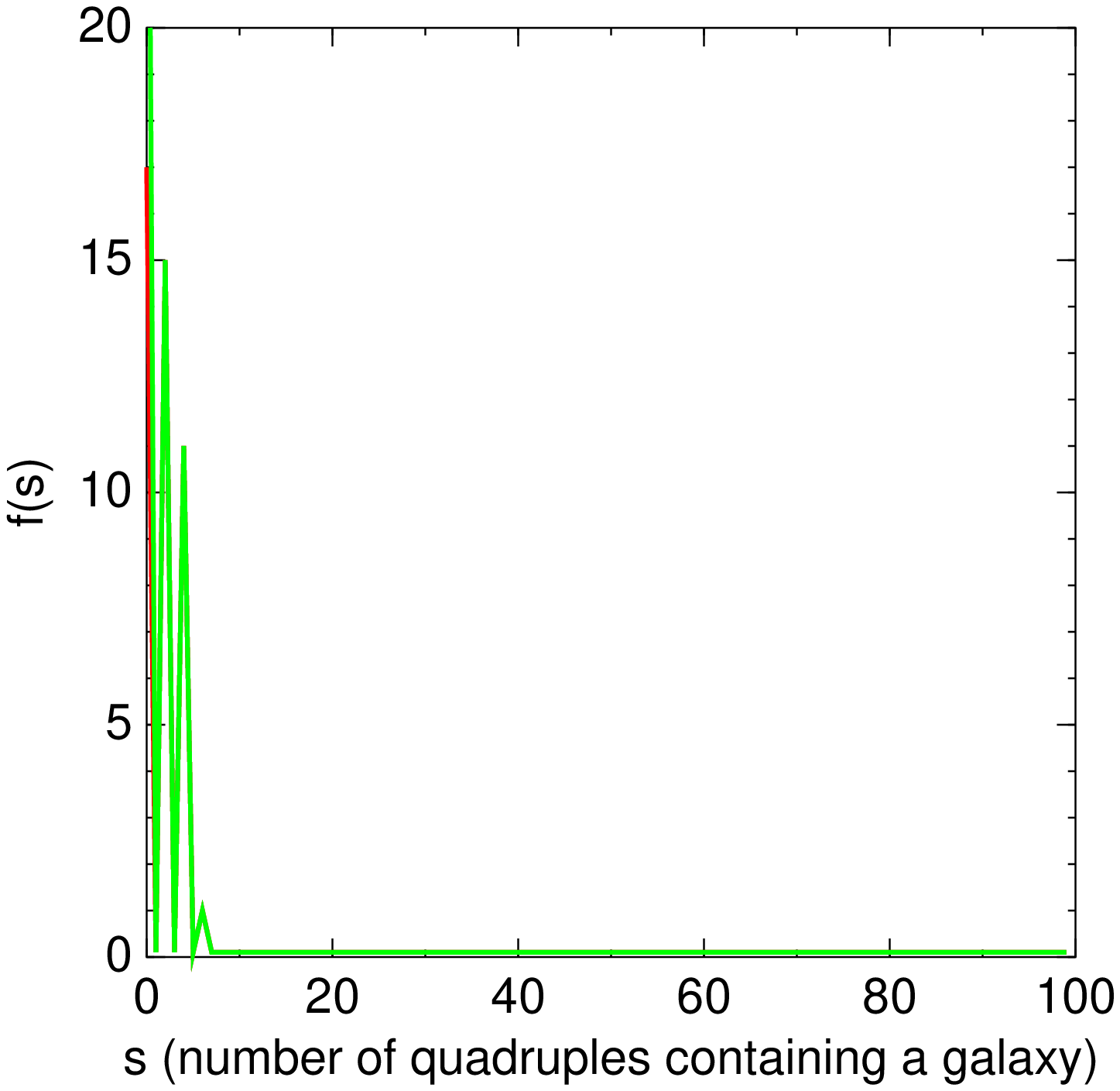} 
    \caption[]{ \mycaptionfont {Quadruple frequencies $N(s)$,
        as for Fig.~\protect\ref{f-quadruples}, but
        a realisation of Gaussian
        uncertainties in the $T^3$ parameters is applied, 
        destroying the topological signal, since none
        of the implied topological images fall in the
        southern field in this realisation.
        \label{f-quadruples-T3-perturb}
    }}
  \end{figure} 
}

\newcommand\fquadrupleseuclidphotz{
  \begin{figure}
    \centering 
    \includegraphics[width=0.8\columnwidth]{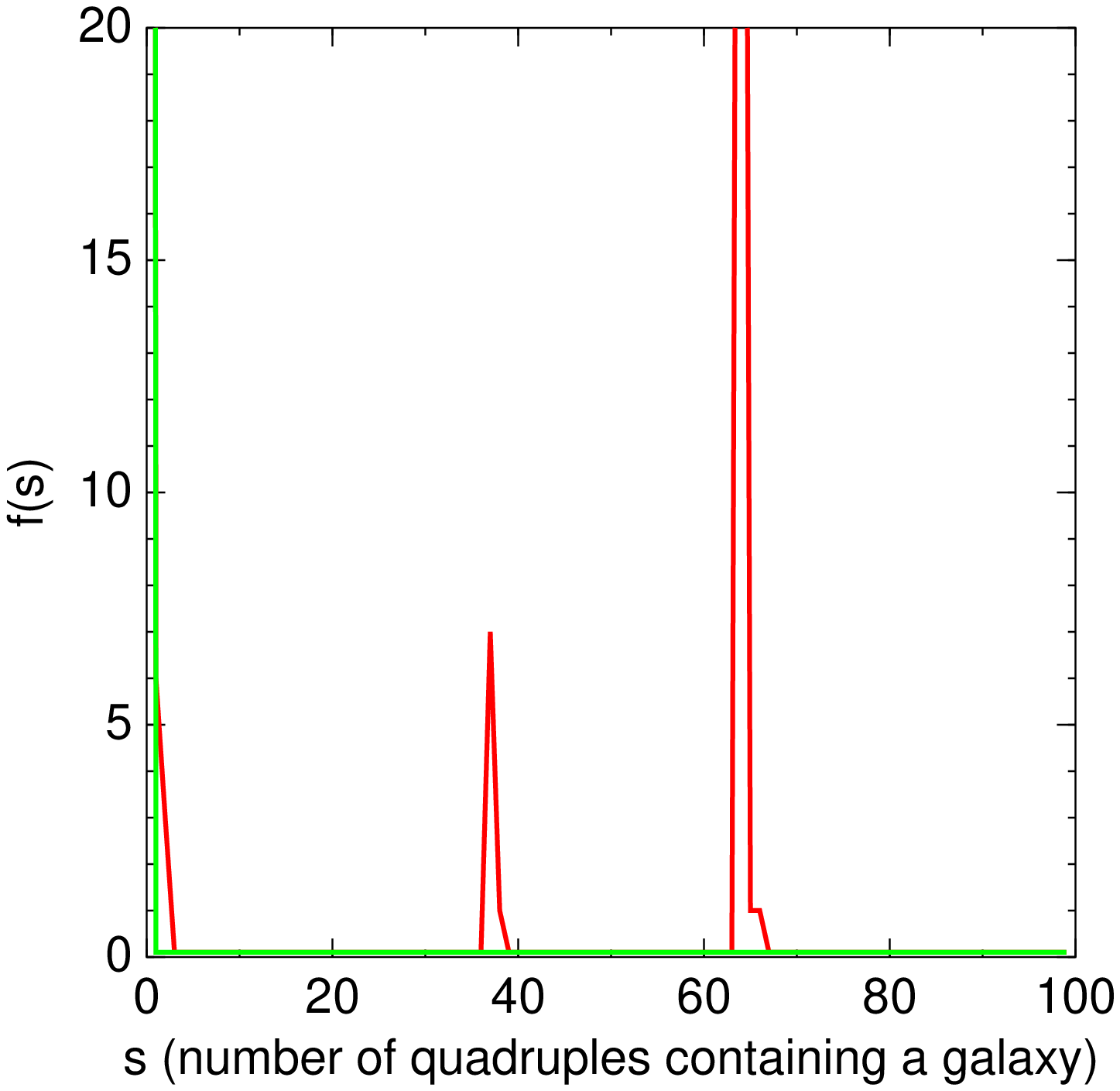} 
    \includegraphics[width=0.8\columnwidth]{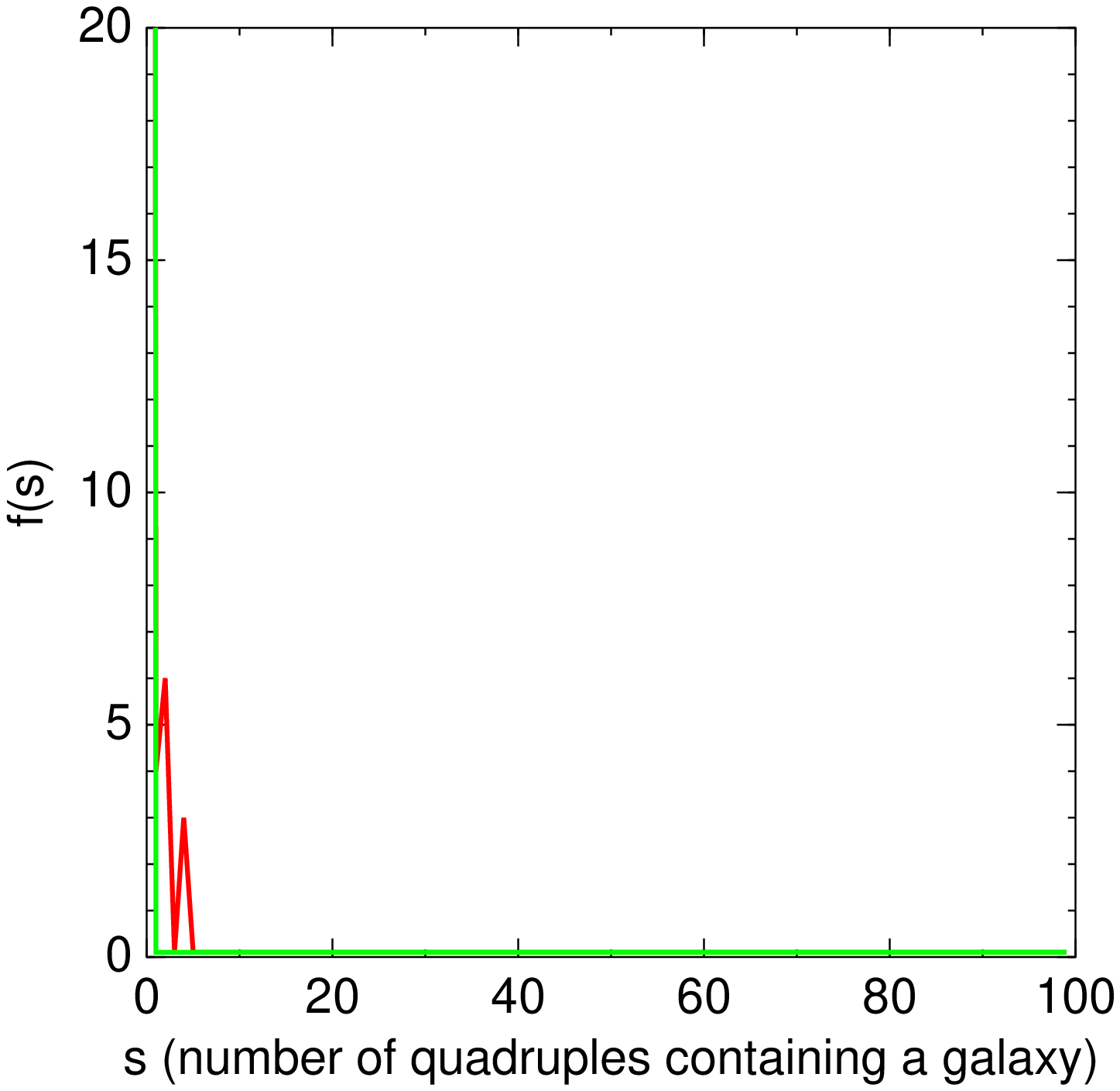} 
    \caption[]{ \mycaptionfont {As for
        Fig.~\protect\ref{f-quadruples}, for fully simulated, multiply
        connected (dark, red online) and simply connected (pale, green
        online) catalogues, over northern and southern fields each of 
        $\approx 64$~deg$^2$ centred at antipodal matched disc centres,
        with $4.3 < z < 6.6$,
        and successive filter parameters
        $\epsilon = 2.0 {\hMpc}, 
        \Delta t = 1.0 {\hGyr},
        \epsBoP = 1 {\hMpc}$, 
        and \prerefereechanges{(in the simply connected case)}
        an input auto-correlation function
        $r_0 = 10.0{\hMpc}, \gamma = 1.8, r_{\min} = 1.0{\hMpc}, r_{\max} = 100.0{\hMpc}$
        [Eq.~(\protect\ref{e-effective-xi})].
        A total of 300 simulated
        objects (north plus south) are present in each catalogue.
        {\em Top:} Spectroscopic redshifts.
        {\em Bottom:} 
        Photometric redshifts with
        $\sigma(\betapec) = 0.005$, 
        i.e. $\sigma(\Delta z) \approx 0.03$.
        \label{f-quadruples-euclid-photz}
    }}
  \end{figure} 
}

\newcommand\fconflevel{
  \begin{figure}
    \centering 
    \includegraphics[width=1.0\columnwidth]{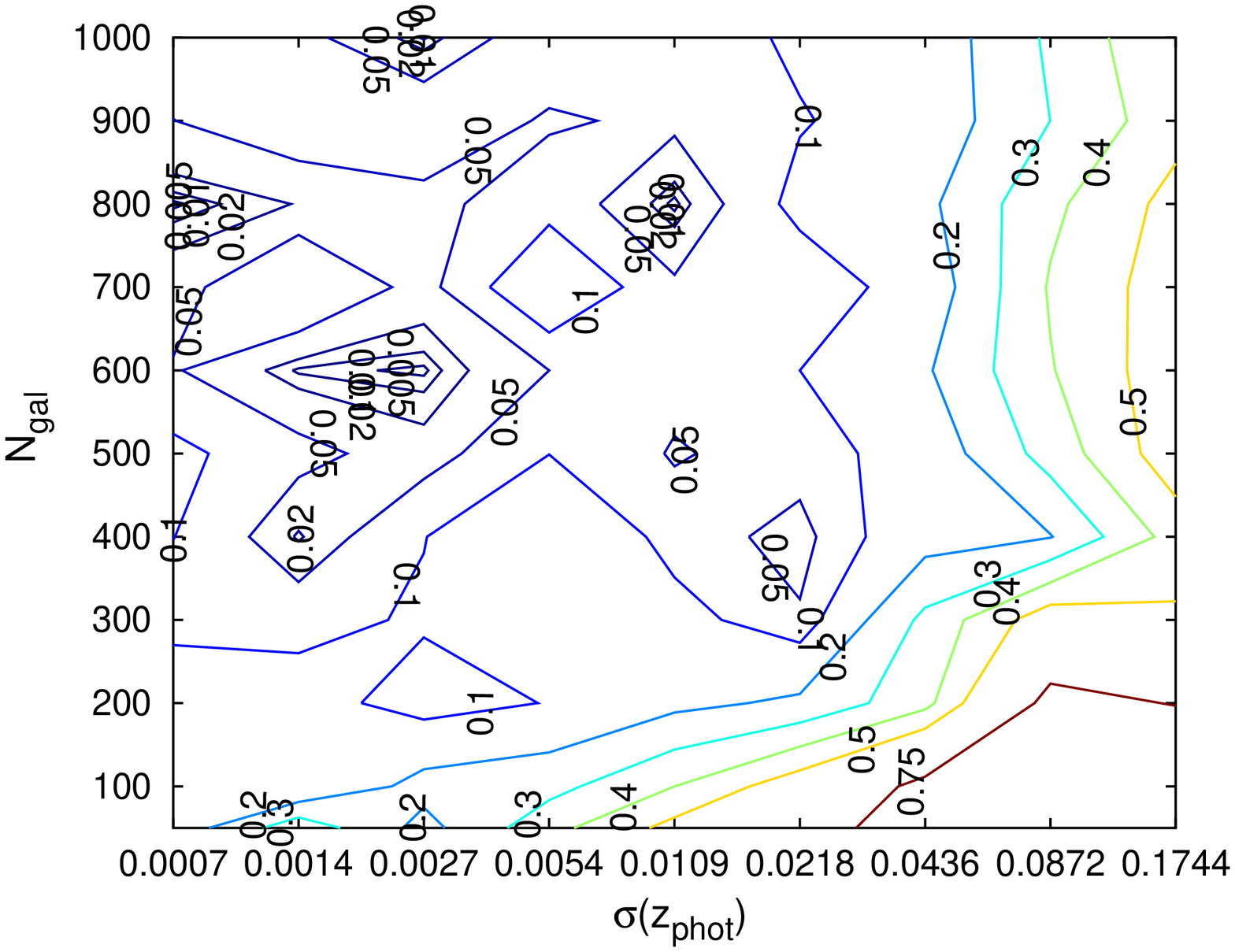} 
    \includegraphics[width=1.0\columnwidth]{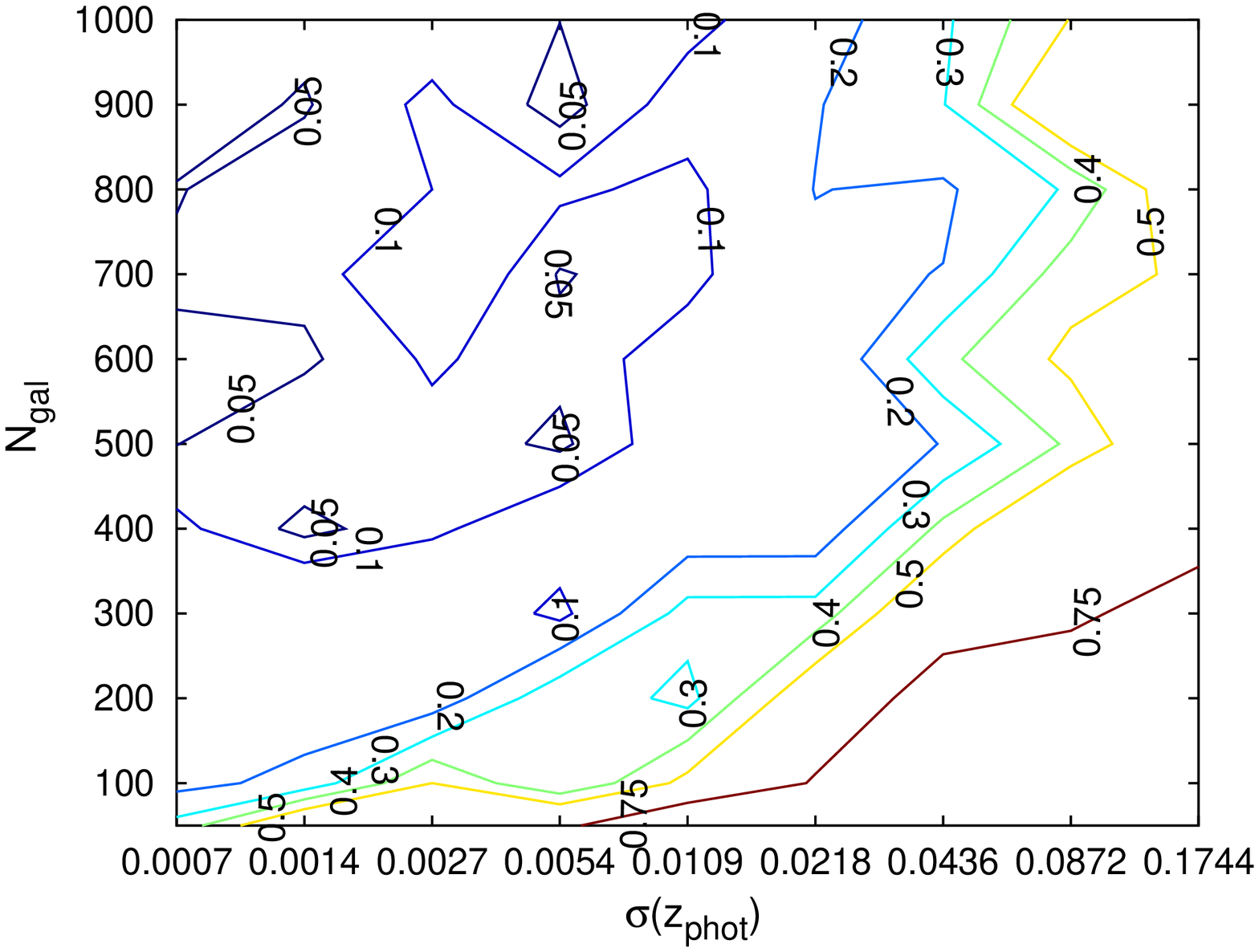} 
    \caption[]{ \mycaptionfont {Expectation values $\probbetabest$ of
        falsely rejecting (or falsely accepting) the $T^3$ hypothesis
        for optimal choices of $s^*$
        [Eq.~(\protect\ref{e-defn-probbetabest})], as a function of
        photometric redshift $\sigma(\zphot)$ [estimated as
          $(1+(z_1+z_2)/2) \sigma(\betapec)$, using $\sigma(\betapec)
          \ll 1$ and Eq.~(\protect\ref{e-z-mult-simpler}), where $z_1
          < z < z_2$ is the redshift range], and the number of
        galaxies $N_{\mathrm{gal}}$ per combined (north plus south)
        catalogue.  The lowest probabilities $\probbetabest$ (highest
        significance) results are for low $\sigma(\zphot)$ and high
        $N_{\mathrm{gal}}$, at top-left in each of the panels.
        {\em Top:} 
        $4.3 < z < 6.6$, solid angle per survey direction (south or north)
        $\omega \approx 100$~deg$^2$, $\Delta t = 1$~{\hGyr}.
        The other parameters are identical to those for
        Fig.~\protect\ref{f-quadruples-euclid-photz}. 
        {\em Bottom:} Same, except that $\Delta t = 0.1$~{\hGyr}.
        \label{f-conf-level}
    }}
  \end{figure} 
}

\newcommand\fconflevelzfourzseven{
  \begin{figure}
    \centering 
    \includegraphics[width=1.0\columnwidth]{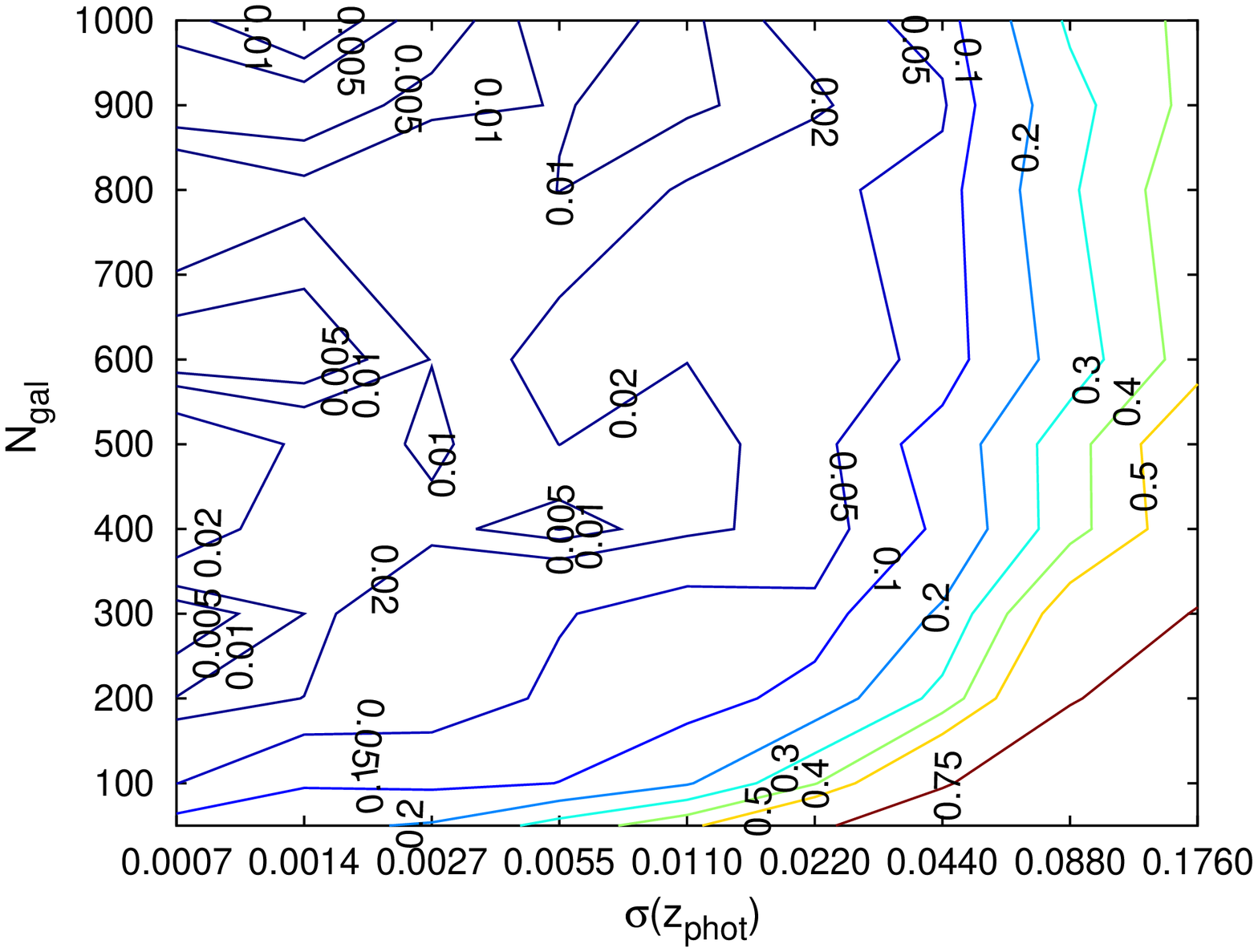}
    \includegraphics[width=1.0\columnwidth]{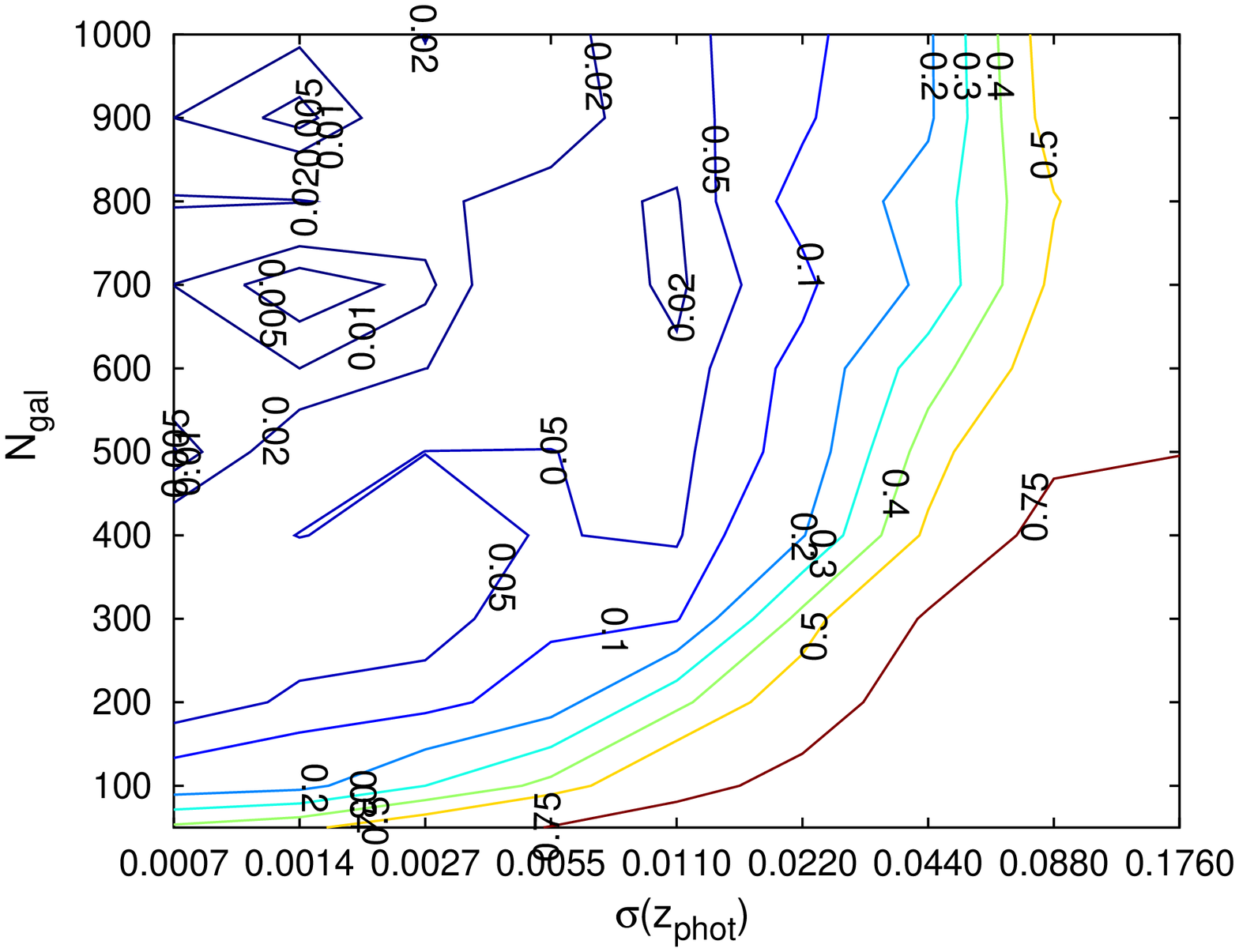}
    \caption[]{ \mycaptionfont {Expectation values $\probbetabest$ of
        falsely rejecting (or falsely accepting) the $T^3$ hypothesis,
        as for Fig.~\protect\ref{f-conf-level}, but for a thicker,
        wider survey, over $4 < z < 7$ and $\omega \approx 196$~deg$^2$.
        {\em Top:} $\Delta t = 1$~{\hGyr},
        {\em Bottom:} $\Delta t = 0.1$~{\hGyr},
        \label{f-conf-level-zfourzseven}
    }}
  \end{figure} 
}

\newcommand\fbench{
  \begin{figure}
    \centering 
    \includegraphics[width=1.0\columnwidth]{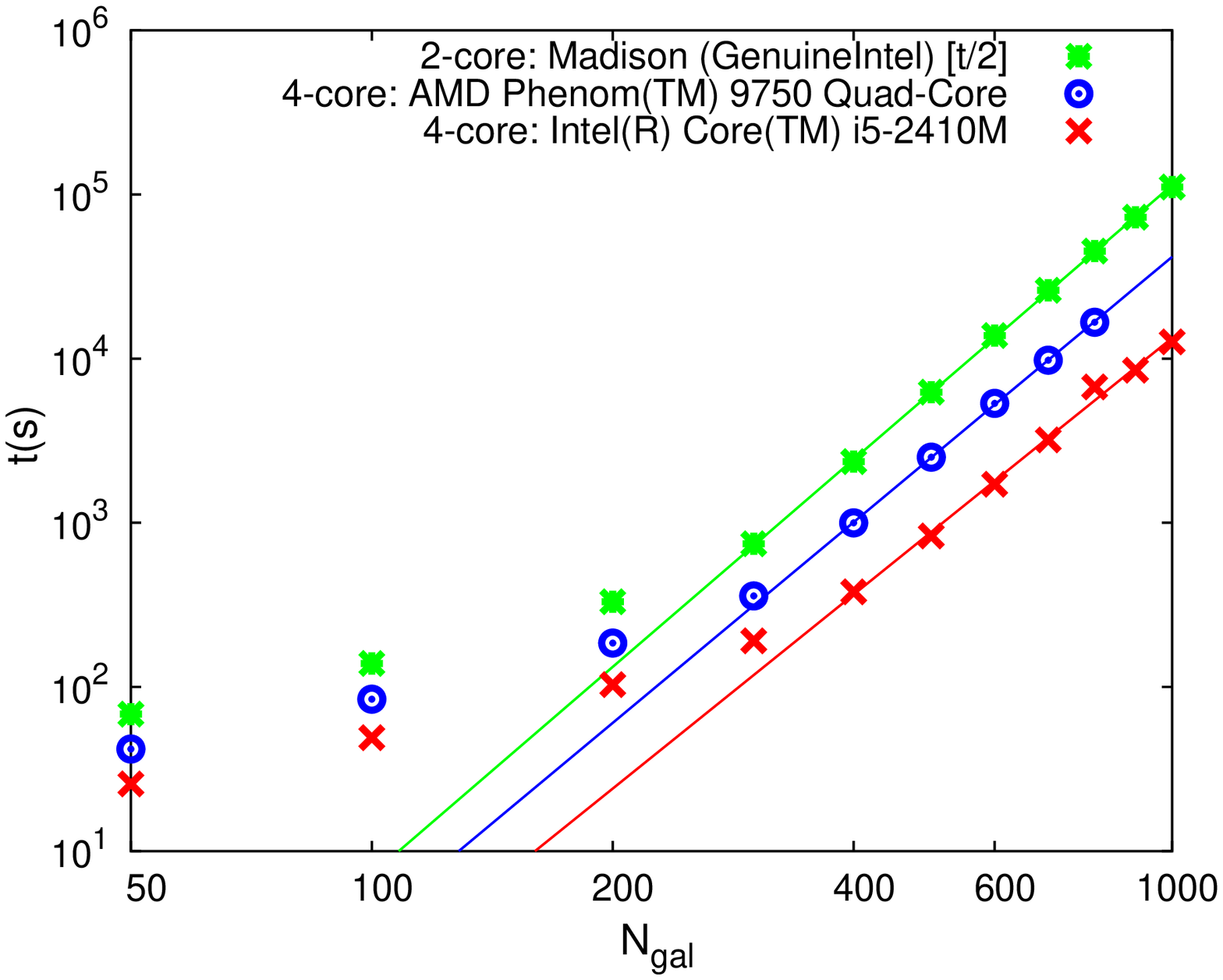}
    \caption[]{ \mycaptionfont {Confidence level calculation time in
        seconds for quadruple searches of (the equivalent of) 30 realisations
        for a single parameter set, as a function of the total
        number of galaxies $N_{\mathrm{gal}}$, for the simulations shown in
        Figs~\protect\ref{f-conf-level} and 
        \protect\ref{f-conf-level-zfourzseven}, for 
        \prerefereechanges{three} different
        processor sets, as labelled.
        The times for the first \prerefereechanges{processor set are 
          halved to} give the equivalent of four cores.
        Power-law fits \prerefereechanges{to $N_{\mathrm{gal}} \ge 400$}
        are shown.
        \label{f-bench}
    }}
  \end{figure} 
}

\newcommand\fdeltartheta{
\begin{figure}
  \centering 
  \includegraphics[width=1.0\columnwidth]{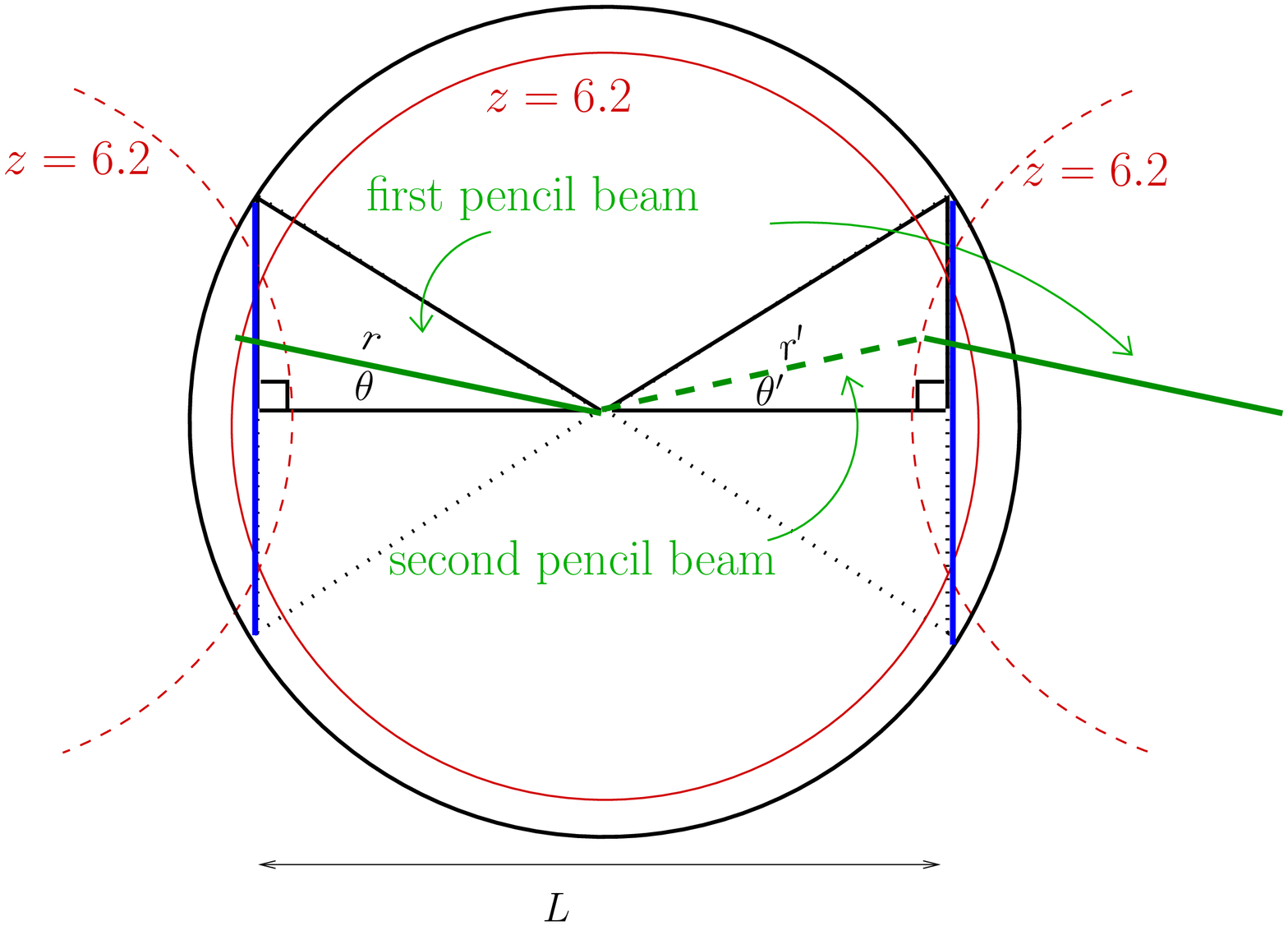} 
  \caption[]{ \mycaptionfont 
    \protect\postrefereechanges{{Relation between topological image
      pair with respect to observer and matched disc centres, 
      as in Fig.~\protect\ref{f-matched-beams}. The object is at
      comoving radial distances $r$ and $r'$, respectively,
      and at angles $\theta$ and $\theta'$, respectively, from
      the matched disc centres. The relation between
      these is given in Eq.~(\protect\ref{e-delta-r-theta}).
      \label{f-delta-r-theta}
  }}}
\end{figure} 
}


\section{Introduction}  \label{s-intro}
The predictions implied by cosmic topology interpretations of the lack of large-scale
(hereafter, $r \gtapprox 10{\hGpc}$)
structure in the cosmic microwave background (CMB) maps
are relevant for the design of 
deep redshift observational strategies.
The large-scale structures in the CMB
as observed by the COsmic Background Explorer (COBE) and the Wilkinson
Microwave Anisotropy Probe (WMAP) \citep{WMAPbasicMNRAS},
i.e., in particular, the second moment of the temperature fluctuation distribution,
can be statistically analysed 
using either (i) spherical \prerefereechanges{harmonics} analysis of the temperature
fluctuations, which 
\postrefereechanges{at large angular scales yields estimates that depend
  strongly on statistical assumptions 
  if observations near the galactic plane
  are contaminated
  \citep[][and references therein]{Copi09},}
 or (ii) the angular or spatial
two-point auto-correlation function of the temperature
fluctuations. 
\postrefereechanges{[In principle, all 
orders of the correlation functions are taken into account in integrated
form by using Minkowski functionals \citep{MeckeBW94};
for iso-temperature excursion sets in relation to CMB analysis 
see \citet{SchmalzBuch97,SchmalzGor98,Ducout12}.]}
Interpreting the (second moment) large-scale harmonics 
to be generated by statistically independent Gaussian distributions 
requires an unlikely ``conspiracy'' between the harmonics at low $l$ values
\citep{Copi07,Copi09,Copi10,Copi10b}
in order to match 
the nearly zero value of the (two-point) auto-correlation function 
at large ($r \gtapprox \rSLS \approx 10.0{\hGpc}$) pair separations
(\citealt[Sect.~7, Fig.~16][]{WMAPSpergel};
\citealt[also Fig.~1][]{RBSG08})
where $\rSLS$ is the comoving distance to the surface of last
scattering (SLS). Thus, 
\postrefereechanges{applying
Occam's razor, our interpretation is} that an approximately zero
large-scale auto-correlation function provides a simpler statistical
model for the largest scales than the harmonic analysis.

The Friedmann-Lema\^{\i}tre-Robertson-Walker (FLRW) models of the
Universe assume that the comoving spatial
section is a 3-manifold, having both curvature and topology
\citep[e.g.,
][]{deSitt17,Fried23,Fried24,Lemaitre31ell,Rob35}.\footnote{See
  \protect\citet{Lemaitre31elltrans} for an English translation of
  \protect\citet{Lemaitre31ell}, with Lema\^{\i}tre's 
  observational estimate 
  $H_0 \approx 600$km/s/Mpc
  excluded
  \protect\postrefereechanges{\protect\citep[e.g. ][]{vdB11Lemaitre,Block11Lemaitre,Luminet13Lemaitre}}.}  
For recent reviews and terminology such as ``fundamental domain'',
``covering space'', and ``topological lensing'' (where the topology of
a spatial section of the Universe can be thought of as a ``lens'', in 
analogy with gravitational lensing), see 
\citet[e.g.,][]{LaLu95,Lum98,Stark98,LR99,RG04}, or shorter, 
\prerefereechanges{\citet{Rouk00BASI,Lum06Brazil}.} 

Structures
bigger than the fundamental domain of an FLRW model 
(i.e. of the 3-manifold of the comoving spatial section)
cannot physically
exist, so observations that are most easily interpreted in the
universal covering space (apparent space) should approximately reveal
this (\citealt{Star93,Stevens93}; though see also Fig.~1,
\citealt{Rouk96}).  Thus, 
\postrefereechanges{according to several analyses,}
the simpler interpretation of the
large-scale CMB observations is that the Universe is a multiply
connected FLRW model rather than a simply connected FLRW model: the
near-zero large-scale two-point 
auto-correlation reveals the size of the Universe. Among the
models that are considered to provide better fits to the WMAP data
than the infinite flat model are the Poincar\'e dodecahedral space
($S^3/I^*$, \citealt{LumNat03,Aurich2005a,Aurich2005b,Gundermann2005,Caillerie07,RBSG08,RBG08}),
\prerefereechanges{the Picard space 
($H^3/\mathrm{PSL}_2(\mathbb{Z}[i])$, \citealt{Picard1884,AurichPicard04})},
the regular 3-torus 
($T^3$, \citealt{WMAPSpergel,Aurich07align,Aurich08a,Aurich08b,Aurich09a}),
and the ``2-torus'' ($S^1 \times S^1 \times \mathbb{R}$, \citealt{AslanMan11}),
though some analyses, also assuming an FLRW (homogeneous) metric,
favour the infinite flat model \citep{KeyCSS06,NJ07,BielB11}.

While there is a lack of consensus in fitting models to the WMAP data,
the $T^3$ model fits have a particular advantage from an observational
point of view: they imply sub-SLS topological lensing at redshifts
that are becoming observationally realistic. Many cosmic topology CMB analyses
consider ensembles of possible sizes and orientations of the
fundamental domain of a given model. A more empirical approach 
requires that the fundamental domain of the
real Universe must have a specific size and orientation in
astronomical coordinates. We live in a specific realisation of the physical
processes that led to our Universe, not an ensemble of realisations.
 \citet{Aurich08a} applied the optimal
cross-correlation method \citep{RBSG08,RBG08} to find a specific
$T^3$ solution.
This solution implies
equal-redshift topological lensing at redshifts $z \gtapprox 6$, the
redshift range at which many detections are expected with the
upcoming generation of new instruments and telescopes. Thus, this
candidate 3-manifold is, in principle, empirically testable independently of
the CMB.

\tTthrlb

\citet{RK11} presented a corollary of the matched circles principle
\citep{Corn96,Corn98b}: the {\em matched discs} principle. In general,
topologically lensed pairs of images of a given physical object occur at
different redshifts. This complicates observational tests: the
lifetimes of quasars are short compared to typical differences in
lookback times, and at high redshifts, a galaxy seen in one direction
at a redshift of $z_1$ may be absent at an expected position with a higher redshift $z_2 > z_1$
because the initial starburst occurs after $z_2$ and before $z_1$.
Matched discs minimise this problem, by selecting the set of spatial
positions---discs---for which a topologically lensed pair of any given
galaxy occurs at identical redshifts in the two matched discs (Fig.~1,
\citealt{RK11}). The redshift is lowest at the centre of a disc and
increases radially outwards.  The Poincar\'e dodecahedral space
matched discs occur with $z_{\min} = 106\pm18$, so detection of
gravitationally collapsed objects (galaxies) would require the
existence and detection of rare high overdensity peaks of the density fluctuation
distribution that collapsed very early.
In contrast, the \citet{Aurich08a} $T^3$ candidate has 
$z_{\min} \sim 5$ (Fig.~10, Sect.~3, \citealt{Aurich08a}).
While relatively small numbers of quasars have been detected
with $z \gtapprox 5$, many Lyman break galaxies (LBGs) and Lyman alpha
emitters (LAEs) in this redshift range have now been observed
[with spectroscopic confirmations up to $z=8.6$ \citep{Lehnert8p6z2010}]
and many more are likely to be detected with the new instruments
and telescopes of the coming decade. LBGs and LAEs have a
topological lensing advantage over quasars: the emission from 
a given object is expected to be much more isotropic than 
the beamed jets of a quasar, and the lifetime is likely to 
be much longer, because of long galaxy dynamical time scales and stellar
lifetimes.

Thus, the \citet{Aurich08a} solution is going to be testable over
matched discs of many square degrees using observations to be made
over the next few years
(e.g., VISTA/VIKING, Subaru/CISCO, EUCLID, VLT/X-Shooter, VLT/MUSE, JWST; 
see \SSS\ref{s-instruments} for details).
Although surveys that are approximately
limited to a narrow band in redshift correspond to thin shells rather
than thin discs, surveys over solid angles $\ll 4\pi$ (e.g. many
square degrees) can strongly overlap with matched discs if made over
the predicted redshift ranges.
Since observational and theoretical uncertainties in the candidate 3-manifold
cannot be avoided, a successive
filter method for finding sets of likely topologically lensed 
pairs \postrefereechanges{\citep{Rouk96,ULL99a,MareckiRB05,FujiiY11a,FujiiY13}} is preferable to 
the pair separation histogram (PSH) method \citep{LLL96}, as the former
can reveal very weak topological lensing signals \citep{FujiiY11a}.
Here, a \prerefereechanges{successive} filter method, motivated by the 
matched discs principle, is applied to 
existing observational data and planned or possible surveys,
using numerical simulations,
in order to investigate observational
strategies that can reject the $T^3$ solution, on the assumption
that the FLRW metric is close enough to physical reality.

In \SSS\ref{s-method-T3-WMAP}, we briefly present the \citet{Aurich08a}
$T^3$ solution, matched discs, and not-quite-matched beams.
Existing and planned or
possible observations that can be used to predict sky positions and
redshifts of topologically lensed copies are presented in
\SSS\ref{s-method-Subaru-etc}.  For existing observations, 
we apply the successive
filter method to two partly simulated catalogues.
The signal is represented by analysing the union of 
the observational catalogue with
its topologically lensed images 
on (approximately) the opposite side(s) of the sky.
Noise is represented by analysing the union \prerefereechanges{of} 
the observational catalogue 
with \prerefereechanges{a simulated} (non-lensed)
data set in the region of comoving space 
where lensed objects would be found.
Thus, signal is compared to noise. 
For planned or possible observations,
the data are fully simulated, comparing a multiply
connected $T^3$ simulation with a simply connected
$\mathbb{R}^3$ simulation.
The simulation methods are described in \SSS\ref{s-method-naive}, and the successive filter method 
is presented in \SSS\ref{s-method-filters}.
Statistical tests, either to find a low probability 
of falsely excluding the $T^3$ hypothesis, 
or to find a low probability of falsely detecting evidence
in favour of the hypothesis, are presented 
in \SSS\ref{s-method-beta-alpha}.

\fxiilcmain

\fmatchedbeams

\fdeltartheta

Results are presented in \SSS\ref{s-results} and
conclusions are given in \SSS\ref{s-conclu}.  All distances are {FLRW}
comoving distances
except if stated otherwise, and 
$\Omega_m$ and $\Omega_\Lambda$ are the
dimensionless matter density and dark energy density, respectively. 
The Hubble constant is written $H_0 = 100
h$~km/s/Mpc and $c$ is the conversion factor from time units to space
units \citep[e.g.][]{TaylorWheeler92}.

\section{Method} \label{s-method}

\subsection{The \protect\citet{Aurich08a} $T^3$ solution, matched 
discs, and not-quite-matched beams}
\label{s-method-T3-WMAP}

\tzmin

The coordinates of the \citet{Aurich08a} $T^3$ solution are
given in Table~\ref{t-T3-lb}. We check the cross-correlation
$\ximc$ \citep{RBSG08} on the 9-year WMAP internal linear combination (ILC)
map\footnote{\href{http://lambda.gsfc.nasa.gov/data/map/dr5/dfp/ilc/wmap_ilc_7yr_v4.fits}{{\tt http://lambda.gsfc.nasa.gov/data/map/dr5/dfp/}}
\href{http://lambda.gsfc.nasa.gov/data/map/dr5/dfp/ilc/wmap_ilc_9yr_v5.fits}{{\tt ilc/wmap\_ilc\_9yr\_v5.fits}}},
using the corresponding KQ85 galactic contamination mask \postrefereechangesbis{\citep{Bennett13WMAP9MNRAS}}, 
\footnote{The 9-year version of the KQ85 mask \protect\href{http://lambda.gsfc.nasa.gov/data/map/dr5/ancillary/masks/wmap_temperature_kq85_analysis_mask_r9_9yr_v5.fits}{{\tt http://lambda.gsfc.nasa.gov/data/map/dr5/ancillary/masks/}}
\protect\href{http://lambda.gsfc.nasa.gov/data/map/dr5/ancillary/masks/wmap_temperature_kq85_analysis_mask_r9_9yr_v5.fits}{{\tt wmap\_temperature\_kq85\_analysis\_mask\_r9\_9yr\_v5.fits}} hides about 25.2\% of the sky.}
and show the dependence of the
sub-gigaparsec-scale mean cross-correlation 
\begin{equation}
\prerefereechanges{\left<\ximc\right>}_{r}(\prerefereechanges{L}) := \frac{1}{r} 
  \int_0^{r} \ximc(r',\prerefereechanges{L}) \mathrm{d}r' 
  \label{e-defn-xim}
\end{equation}
in Fig.~\ref{f-xi-ilc-85}. 
The cross-correlation $\ximc$ is a two-point correlation function
that at small scales correlates pairs of points that are observed
to be separated by large distances but according to the hypothesised
3-manifold are also separated by a small distance. The cross-correlation
should be low if the hypothesis is wrong \citep[Fig.~2][]{RBSG08}
and high if the hypothesis is correct \citep[Fig.~3][]{RBSG08}.
The strongest cross-correlation in Fig.~\ref{f-xi-ilc-85} occurs for
$\LTthr \approx 3.80 \pm 0.05$ (in units of $c/H_0$), where the
uncertainty is the half-width of the maximum at its base. The
\citet{Aurich08a} uncertainties are estimated from the
pseudo-probability function used in the Monte Carlo Markov Chain
search for an optimal solution. Since the pseudo-probability estimator
is not a true probability, other methods are needed to estimate a
realistic uncertainty. Figure~9 of \citet{Aurich08a}, showing the
estimates in different foreground-subtracted single waveband WMAP
maps, suggests $\sigma(\LTthr) \approx 0.02$, combining random and
some systematic sources of error (ILC versus single band maps).  The
shift by $0.05 (c/H_0)$ between \citet{Aurich08a}'s estimate of
$\LTthr$ for WMAP5 data and that shown in Fig.~\ref{f-xi-ilc-85}
suggests that inclusion of systematic error would give
$\sigma(\LTthr) \gtapprox 0.05$, i.e. 150{\hMpc},
rather than $\sigma(\LTthr) = 0.02$. The angular 
uncertainty given by \citet{Aurich08a}, $2^\circ$, is similar
to that of the Poincar\'e dodecahedral space solution found
earlier with the optimal cross-correlation method
\citep{RBSG08,RBG08}, i.e. a few hundred comoving {\hMpc} in 
a tangential direction at a radial distance of 5 to 10 {\hGpc}.

Leaving aside these uncertainties for the moment, the observationally
most dramatic element of this $T^3$ solution is revealed by 
considering the redshift of an object topologically lensed in 
opposite directions along a fundamental axis. \citet{Aurich08a} 
shows this in his Fig.~10 and briefly discusses it. Here, we use
the principle of matched discs \citep{RK11}, shown with some extra
detail in Fig.~\ref{f-matched-beams}, and Table~\ref{t-zmin}, listing
the redshifts at the centres of a pair of matched discs and at
circles increasing radially outwards to form the pair of matched discs.
The fractional sky coverage $\omega/(4\pi) = 12 \pi(1 - \cos\beta )/(4\pi)$,
for discs of angular radius $\beta$,
is also shown. 

Given the rapidly increasing numbers of observed
objects in the $5 \ltapprox z \ltapprox 6$ range, the beginning
of detections around $6 \ltapprox z \ltapprox 8$, and the large
fractional covering of the sky by the matched discs, it is 
obvious that independently of the few hundred {\hMpc} uncertainties 
in the $T^3$ solution, the observational data to reject or confirm
the solution to very high statistical significance are going to
become available over the next decade or two. The absence of 
topologically lensed images of quasars can be interpreted as 
a problem of their short lifetimes and highly anisotropic nature
(beamed jets), but the absence of topologically lensed images of
early forming galaxies would be much more difficult to explain.

Let us return to the uncertainty in the $T^3$ solution parameters.
Pencil beam observations (``deep fields'') that probe to $z \sim 6$
have typical widths of at most a few arc\-minutes, i.e. a few {\hMpc} in
comoving thickness at these redshifts.  Thus, an accuracy of a few
hundred {\hMpc} suggests that deep fields of a few degrees in
size---pointed in the appropriate directions---will most likely be needed.
The fourth row of entries in Table~\ref{t-zmin} shows the redshifts at the
$\beta=10^\circ$ radius circle for $\LTthr=4.0$, i.e. $4\sigma$
greater than the WMAP9-ILC estimate shown in 
Fig.~\ref{f-xi-ilc-85} if $\sigma$ is defined as the 
absolute difference
between the \citet{Aurich08a} WMAP5 estimate and the WMAP9-ILC 
estimate made here. Thus, 
for surveys limited to, e.g. $z < 6.2$,
observations closer to the face centres
would better test the solution.

Figure~\ref{f-matched-beams} also shows a pair of not-quite-matched
beams.  These are not truly matched, since they exactly match only
where they intersect the matched discs. The topologically 
matched parts of the two beams to the left (in Fig.~\ref{f-matched-beams})
of their respective discs occur in the left-hand copy of the beam 
at slightly higher
redshifts and in the right-hand copy at slightly lower redshifts, with
respect to the redshift of the intersections with the matched discs.
Since realistic surveys usually have wide redshift distributions, there
is a fair chance of covering the topologically lensed region.
However, this requires that the survey
of the right-hand copy (in this figure) of the beam be wide enough in
solid angle to cover the projection of the ``beam'' onto the sky, since
the right-hand copy is not a beam from the observer's point of view.
For a small angular distance of the left-hand copy of the beam from
the matched disc centre, the projected solid angle of a small portion
of the right-hand copy will not be too large.
Thus, for a small angular offset from the matched disc centres (axes
of the fundamental domain), ``not-quite-matched beams'' can potentially
test topological lensing hypotheses.
\postrefereechanges{The relation between an object near a matched disc centre
  and its topological image is illustrated (for the $T^3$ case) in 
  Fig.~\ref{f-delta-r-theta}, giving
  \begin{eqnarray}
    \theta' &=& \mathrm{atan} 
    \left(\frac{r \sin \theta}{L - r\cos \theta} \right) \nonumber \\
    r' &= & \frac{r \sin \theta}{\sin \theta'}.
    \label{e-delta-r-theta}
  \end{eqnarray}}

While matched discs and not-quite-matched beams indicate the parts of
the sky that should be observed, detecting a significant statistical signal
and generating a list of candidate pairs of topologically lensed objects
requires a sensitive statistical method. This is presented
below in \SSS\ref{s-method-filters}. However, first we need to consider
existing, planned and possible surveys in the three-dimensional 
regions of interest.

\tTthrsofar

\tdeepfields

\tsubaru

\subsection{Observational catalogues}
\label{s-method-Subaru-etc} 

\subsubsection{Existing observations}
Table~\ref{t-T3-so-far} shows that almost all of the $z \sim 6$
objects that are so far known near the $T^3$ axes are those near the
high galactic latitude, northern fundamental direction 1, although a few 
are also known near the corresponding southern fundamental direction 4.
Most of the northern objects are from the Subaru Deep Field
observed with the Cooled Infrared Spectrograph and Camera for [the] OH airglow suppressor (CISCO, \postrefereechangesbis{\citealt{MotoharaSubaruCISCO02}}) on 
the 8.2~m Subaru telescope
\citep{SubaruDF01MNRAS,SubaruDeep2006},
listed along with some of the other well-known deep fields in Table~\ref{t-deepfields}.
The galactic coordinates and redshifts of the northern objects are listed 
as part of Table~\ref{t-subaru}.

\subsubsection{Planned and possible observations} \label{s-instruments}

First let us consider high-redshift quasars.
The VISTA/VIKING 4-m class telescope project should discover quasars
at $z \sim 7$ over about 1500 deg$^2$ centred on the northern and
southern Galactic poles\footnote{The criterion ``centred on the
  northern Galactic cap'' presumably means ``a large solid angle
  centred on the northern Galactic pole''.}  \citep{VISTAFindlay11}.
Although Table~\ref{t-zmin} here and the expected completeness levels
shown in Fig.~13 of \citet{VISTAFindlay11} indicate that coverage of
the desired regions is possible (depending on the value of $\Omm$),
only about 8 quasars are expected in the whole survey ({\SSS}4.1,
\citealt{VISTAFindlay11}), clearly too low to obtain significant
evidence either for or against the $T^3$ candidate.

\citet{Polsterer12} find 22,992 photometric quasar candidates with $5.5 < z < 6.2$
and an estimated error of $\sigma(\Delta z) = 0.087$, expecting about half the
candidates to be true quasars. However, the solid angle from which the quasars 
are selected is about $2\pi$, giving a
candidate surface density of about 1/deg$^2$
and an expected spectroscopic number density 
of about 0.5/deg$^2$ if a complete spectroscopic
followup were to be performed. This is about 100 times higher than for the VISTA/VIKING
survey (as presently designed), but still somewhat low. While quasars 
dominated high-redshift records for the decades when $z > 1$ was considered a high redshift
for a spectroscopically confirmed extragalactic object, this seems less likely at $z \sim 6$. For a given apparent magnitude limit, 
it is clear that LBG and LAE catalogues are present in much high
number densities at $z \sim 6$.

Plans for LBG and LAE searches typically focus on the existing ``deep fields'', 
with the aims of achieving wide coverage across the electromagnetic spectrum.
Several of these fields and their angular distances $\theta_i$ from the closest $T^3$ axes
are given in Table~\ref{t-deepfields}, indicating which fields would
be the most useful for testing the $T^3$ candidate via LBG and LAE searches.
Comparison with Table~\ref{t-zmin} indicates that for 
$5 \ltapprox z \ltapprox 7$ 
searches, the Subaru Deep Field and its corresponding northern region
is optimal, while for $5.5 \ltapprox 8$ searches, the Virmos Very Deep Survey fields 0226$-$04 and 1400+05 and their northern and southern, respectively,
counterparts would be worth observing.


The EUCLID mission\footnote{\protect\url{http://sci.esa.int/euclid}}
planned for launch in
2019\footnote{\protect\href{http://sci.esa.int/science-e/www/object/index.cfm?fobjectid=49385}{{\tt http://sci.esa.int/science-e/www/object/}}\\
\protect\href{http://sci.esa.int/science-e/www/object/index.cfm?fobjectid=49385}{{\tt index.cfm?fobjectid=49385}}}
has a deep survey 
\citep[Sect.~1.3,][]{EuclidScienceBook2010}, the EUCLID optical and NIR
Deep Imaging Survey, 
planned over 40 deg$^2$.
This should obtain ``thousands'' of likely LBG's
and LAE's with $z > 6$ based on photometric redshifts
\citep[Sect.~14.2,][]{EuclidScienceBook2010}, i.e., $\gtapprox
50$/deg$^2$.  If a few objects had quadruple multiplicities that made
them very likely topological lensing candidates, then
optical/near-infrared spectroscopic followup with an instrument such
as X-Shooter \citep[300 nm $\ltapprox \lambda \ltapprox$ 2400~nm,][]{Xshooter2011MNRAS} on the Very Large Telescope (VLT) on the
southern objects and corresponding spectra of the northern objects
with the Keck or Subaru telescopes should obtain enough rest-frame
spectral energy distribution information to check whether the would-be
topologically lensed pairs of objects resembled each other more than
would be expected for arbitrary pairs of objects at similar redshifts.
The Multi Unit Spectroscopic Explorer
(MUSE)\footnote{\protect\url{http://muse.univ-lyon1.fr/}} on the VLT,
which should be well-tested by the time that the EUCLID data are
available, would be useful for secondary followup to study the spatial
environments of the candidate lensed objects, especially when the
cosmological time difference between the members of a pair is small
compared to a typical galaxy dynamical time scale.

Thus, we should simulate \prerefereechanges{several} square degrees of
observations with parameters consistent with those estimated for
EUCLID. 
\prerefereechanges{In particular, it would be
interesting} to see
if a statistical signal could be obtained from photometric redshifts
alone, prior to spending high amounts of exposure time on highly
sought after telescope/instrument combinations.

The Ultra-Deep Survey of the James Webb Space Telescope 
(JWST)
should detect galaxies at $z \gtapprox 6$, but over only about
10~arcmin$^2$ \citep[Sect.~2, Table~II,][]{JWSTGardner06SRDMNRAS}.
The JWST's planned Deep Wide Survey should find galaxies
over $1 \ltapprox z \ltapprox 6$ over a larger solid angle,
100~arcmin$^2$ \citep[Sect.~3.6, Table~III,][]{JWSTGardner06SRDMNRAS}.
For an initial uncertainty of about two degrees in the $T^3$ axes,
these surveys are clearly too narrow.

\subsection{Simulations}
\label{s-method-naive} 

Simulating searches for \postrefereechanges{matched} quadruples requires comparison of 
observations in a $T^3$ model to those in an $\mathbb{R}^3$ model.
The existing Subaru Deep Field observations are used to provide
an observationally based simulation of both sorts, in which the southern field is generated for the
$T^3$ model by applying the \citet{Aurich08a} $T^3$ holonomy (Table~\ref{t-T3-lb}) to the Subaru galaxies, 
and \prerefereechanges{simulated independently of real observations} 
in the case of the $\mathbb{R}^3$ model.
For a hypothetical observational program aimed at the centres
of both the 1--4 axis expected matched discs, fully simulated
data are needed in both  cases.

The spatial two-point auto-correlation function $\xi(r)$ could, in principle,
lead to excess non-topological chance isometries if the tolerances for
requiring isometry are not tight enough.  For the low number densities
of objects expected, the effect is unlikely to be strong. 
Half 
of a large number of points are first generated uniformly within 
a redshift shell defined by the required redshift limits. Each 
further point is chosen by randomly choosing an existing 
point, and placing the new point in a 3-Euclidean 
direction chosen uniformly from $4\pi$~ster at a comoving 
distance chosen with the probability 
$  P(r,\diffd r) \propto 1 + \xi(r)$, where
\begin{equation}
  \xi(r) = \left\{ 
  \begin{array}{l l}
    \left(\frac{r_0}{r_{\min}}\right)^{\gamma}, & r \le r_{\min} 
     \\
    \left(\frac{r_0}{r}\right)^{\gamma}, & r_{\min} \le r < r_{\max}
     \\
    0, & r \ge r_{\max},
  \end{array} \right.
  \label{e-effective-xi}
\end{equation}
for a correlation length $r_0$ and power law index $\gamma$, and
numerical cutoffs $r_{\min}$ and $r_{\max}$.  Points that do not fall
in the redshift shell are ignored, and new points are generated in the
same way until the required number fill the shell.  Numerical
measurement of the resulting simulated distributions shows that in
practice, this simulates a stronger correlation than
\prerefereechanges{that} required. Thus, the chance of non-topological
isometries is overestimated, i.e. our results are conservative: real
observations are less likely to give a false positive detection.
\prerefereechanges{Moreover, in order to err on the side of possibly
  underestimating the numbers of matching quadruples in the multiply
  connected case, the correlation function is not used when simulating
  this case; uniform distributions are drawn from instead. Again, this
  is conservative: we slightly underestimate the statistical
  significance of the method (\SSS\ref{s-method-beta-alpha}).}

Peculiar velocities and the uncertainties in 
redshift estimation need to be simulated
\citep{Rouk96}. Photometric redshift errors are those of most
interest here: are they small enough to \prerefereechanges{significantly
discriminate} between signal and noise?
These errors 
are all simulated radially 
using a peculiar rapidity $\phipec$ selected from a Gaussian 
distribution of mean zero and standard deviation 
$\;\mathrm{atanh}(\sigma(\betapec))\;$ for a single input parameter,
the peculiar velocity standard deviation $\sigma(\betapec)$,
and using Eq.~(\ref{e-z-mult-simpler}), 
where $\betapec = \tanh \phipec$ and
$\zpec = [(1+\betapec)/(1-\betapec)]^{1/2} - 1.$ Use of
rapidities rather than velocities avoids the unphysical
case of $|\betapec| \ge 1$ for simulating 
photometric redshifts with high $\sigma(\betapec)$.

Before applying the holonomies in a given simulation, 
random Gaussian errors are added to 
the $T^3$ axis parameters given in Table~\ref{t-T3-lb}.
``Observations'' of the simulated catalogues are carried out in
galactic coordinate limited regions in the expected position(s).

\subsection{The successive filter method for type I pairs}
\label{s-method-filters}

\citet{LLU98} introduced the terminology of type II pairs
and type I pairs (or $n$-tuples), that had been introduced earlier by \citet{LLL96}
and \citet{Rouk96}, respectively.  Type I pairs or $n$-tuples can be
thought of as the matching between a local region of objects and its
distant copy. Since a holonomy for an FLRW multiply connected model is
an isometry, the distances among the members within the ``original''
region should correspond to the distances among the members of the
copy of that region.  A Type II pair---in a space such as
$T^3$---corresponds to an object and its copy. In $T^3$, this
separation is a vector in the covering space $\mathbb{R}^3$.
Difficulties in finding statistically significant numbers of either
type of pair given realistic observational parameters and noise led to
a new way of collecting type I pairs \citep{ULL99a} and a successive
filter method \citep{MareckiRB05}.  \citet{FujiiY11a} presented a
variation on the successive filter method, along with an analysis step
that is roughly equivalent 
to collecting together $n$-tuples of mixed type for all $n \ge 4$
or to the \citet{MareckiRB05} bunches of pairs filter
(\SSS~3.2.5, \citealt{FujiiY11a}). The latter step, made possible by
the successive filters, significantly reduces the combinatorial
problem presented in \citet{Rouk96}.  A catalogue containing two
distant regions, each with $N$ objects, might contain just one pair of
matched $n$-tuples across the two regions, but searching for all
$({}^NC_n)^2$ combinations of objects with $N=100$ and $n=7$ would
require $10^{20}$ comparisons of 7-tuples. \citet{FujiiY11a} use toy
simulations to show that a very small number of matched $n$-tuples
can be detected by the full method of successive filtering and
collecting.

The successive filter method implemented here
tests four simulated (and/or real) galaxies at 
$(x,y,z)(i),$ 
$(x,y,z)(j),$ 
$(x,y,z)(k),$ and
$(x,y,z)(l)$ 
via the following filters in the following order,
where $(i,j,k,l)$
is an ordered choice of unequal indices in the list of galaxies.
A crossed quadruple
parameter $\chi(\mathbf{q}) \in \mathbb{Z}_2$ 
is defined for a given
{\em ordered} quadruple $\mathbf{q} = (i,j,k,l)$ in order to relate the
lifetime and bunches of pairs (BoP)
filters \citep[][]{MareckiRB05}. This algorithm
allows the pair separations $d$,
where $d(.,.)$ is the comoving distance between 
two arbitrary points in the covering space $\mathbb{R}^3$, 
the (signed) $x,y,z$ component separations of each pair,
to be calculated first,
and the loop for finding quadruples to be performed over pairs of
pairs. Arithmetically, the change from subtraction
to addition of pair separations in the BoP filter is equivalent
to swapping an $((i,j),(k,l))$ pair of ordered pairs to 
$((i,k),(j,l))$.
The filters are:
\begin{list}{(\arabic{enumi})}{\usecounter{enumi}}
\item
  \prerefereechanges{type I} pair filter: 
  \begin{equation}
    |d(i,j) - d(k,l)| < \epsilon,
    \label{e-pairI-filter}
  \end{equation}
\item
  lifetime filter:
  \begin{eqnarray}
    |t_i -t_k| < \Delta t \; \mathrm{and} \;
    |t_j -t_l| < \Delta t &\Rightarrow &
    \chi((i,j,k,l)) = 0, \; \mathrm{or} \nonumber \\
    |t_i -t_l| < \Delta t \; \mathrm{and} \;
    |t_j -t_k| < \Delta t &\Rightarrow&
    \chi((i,j,k,l)) = 1, \; \mathrm{or} \nonumber \\
    \mbox{otherwise:} && \mbox{filter fails,} \; \nexists\chi
    \label{e-liftime-filter}
  \end{eqnarray}
\item
  BoP filter:
  \begin{eqnarray}
     \Big( \chi((i,j,k,l)) = 0  &\mathrm{and}& 
      \big\vert [x(i)-x(j)] - [x(k)-x(l)] \big\vert  < \epsBoP   \nonumber \\ 
      & \mathrm{and} & \big\vert [y(i)-y(j)] - [y(k)-y(l)] \big\vert < \epsBoP   \nonumber \\ 
      & \mathrm{and} & \big\vert [z(i)-z(j)] - [z(k)-z(l)] \big\vert < \epsBoP \Big)      \nonumber \\ 
      &\mathrm{or}& \nonumber \\ 
      \Big(\chi((i,j,k,l)) = 1 &\mathrm{and}& 
      \big\vert [x(i)-x(j)] + [x(k)-x(l)] \big\vert < \epsBoP   \nonumber \\ 
      &\mathrm{and}& \big\vert [y(i)-y(j)] + [y(k)-y(l)] \big\vert < \epsBoP    \nonumber \\ 
      &\mathrm{and}& \big\vert [z(i)-z(j)] + [z(k)-z(l)] \big\vert < \epsBoP \Big). \nonumber \\
  \label{e-bop-filter}
  \end{eqnarray}
\end{list}
Substituting a difference test for the BoP filter above, i.e.
$|d(i,j) - d(k,l)| < \epsBoP$ and
$|d(i,k) - d(j,l)| < \epsBoP$ for the uncrossed and crossed
pairs of pairs, respectively, would allow non-planar
quadrilaterals (fold a rectangular sheet of paper along its diagonal
to see this), whereas only a parallelogram can represent
a $T^3$ topological quadruple. 

A list of quadruples (pairs of pairs, each associated with a crossed
quadruple parameter $\chi$) that satisfy all three successive filters
is obtained.
An arbitrary galaxy $i$ is a member of $s_i \ge 0$ quadruples.  
The frequency of $s$ in a given simulated set of galaxies is $f(s)$, i.e.
\begin{equation}
  f(s) := | \{ i : s_i = s \}|.
\label{e-defn-Ns}
\end{equation}
If a topological
lensing signal is present, there should be a high number of galaxies
which are a member of many quadruples, $f(s)$ should be
high for high $s$. This is a critical step in the successive filter method,
introduced in Sect.~3.2.5 of \citet{FujiiY11a}: for $s \gg 1$, the
statistic $f(s)$ should be significantly higher in a catalogue
containing topological lensing than in a simply connected catalogue.

\subsection{Statistical significance}
\label{s-method-beta-alpha}
For an arbitrary realisation, 
let us define the cumulative number of galaxies that are each members
of many quadruples
\begin{equation}
  F(s^*) := \sum_{s>s^*} f(s),
  \label{e-defn-F}
\end{equation}
for some quadruplet membership number $s^* > 0$, which 
formalises the word ``many'' in ``many quadruples''.
Thus, for a fixed value $s^*$, 
$F$ is a random variable which we model numerically.
To estimate what observational strategy is required 
to have a low chance of a false negative inference from the data,
i.e. to find $\beta$, the expected probability 
of falsely excluding the $T^3$ hypothesis, 
suppose that the observational result gives $F=i$ for some 
non-negative integer $i$. Then the probability of falsely excluding
the \citet{Aurich08a} $T^3$ hypothesis is the cumulative probability 
$P(F \le i | T^3)$, which
can be estimated numerically by finding the fraction of $T^3$ simulations
for which $F \le i$. 

\postrefereechanges{Since we don't yet know the results of the observations,
we have to weight this 
over 
$p(F=i | \mathbb{R}^3 )$,}
the probability density
function of $i$ for the simply connected case. 
Thus, the expectation value
of the probability of falsely rejecting the hypothesis 
\postrefereechanges{$P(F \le i | T^3)$ (which itself is a random variable)}
is
\begin{eqnarray}
 \probbeta &=& \protect\postrefereechanges{\mathrm{E}[ P(F \le i | T^3) ]} \nonumber \\
 &=& \sum_{i=0}^{\infty} 
 \left( p(F=i | \mathbb{R}^3 ) \; P(F \le i | T^3) \right).
  \label{e-howto-beta}
\end{eqnarray}
Similarly, to estimate what observational strategy is required 
to have a low chance of a false positive inference from the data,
the expected probability of falsely supporting the $T^3$ hypothesis
can be written
\begin{eqnarray}
  && \sum_{i=0}^{\infty} 
  \left( p(F=i | T^3 ) \; P(F \ge i | \mathbb{R}^3) \right). 
  \nonumber \\
  &=&  \sum_{i=0}^{\infty} 
  \left[ p(F=i | T^3 ) \; 
    \left( \sum_{j=i}^{\infty} p(F = j | \mathbb{R}^3) \right)
    \right]. 
  \nonumber \\
  &=&  \sum_{i=0}^{\infty} 
  \sum_{j=i}^{\infty}
  \left[ p(F=i | T^3 ) \; 
    \left(  p(F = j | \mathbb{R}^3) \right)
    \right]. 
  \nonumber \\
  &=&  \sum_{j=0}^{\infty} 
  \sum_{i=0}^{j}
  \left[ p(F=i | T^3 ) \; 
    \left(  p(F = j | \mathbb{R}^3) \right)
    \right]. 
  \nonumber \\
  &=& \probbeta,
  \label{e-alpha-is-beta}
\end{eqnarray}
i.e. the two expectation values are equal.

\fquadruples

\fquadruplesTthreeperturb

For a given simulation parameter set, a set of simulations is
analysed for each $s^* \in \{1, \dots, 10\},$ a range that is found below
(\SSS\ref{s-results}) 
by inspection of $N(s)$ versus $s$ histograms
for typical realisations. The minimum 
\begin{equation}
  \probbetabest := \min_{s^*} \probbeta
  \label{e-defn-probbetabest}
\end{equation}
determines the $s^*$ value for the optimal statistical test for the simulation parameter set,
and the expectation value of the probability 
of falsely excluding the $T^3$ hypothesis or  
of falsely detecting evidence in favour of it.

\section{Results} \label{s-results}

\subsection{Existing observations}
\label{s-results-Subaru-etc} 

Table~\ref{t-subaru} lists observed objects and the implied positions
where topologically lensed images should be observed if the $T^3$
hypothesis is correct, with zero uncertainty in the $T^3$ axis
parameters given in Table~\ref{t-T3-lb}).  Figure~\ref{f-quadruples}
shows the application of the successive filters method under the
assumption that a survey of one square degree in the redshift range $5
< z < 6.2$ is 100\% complete in comparison to the surveys that found
the high galactic latitude sample.  The southern survey field is
centred on the expected median celestial position of the implied
objects (the fourth and fifth columns of
Table~\protect\ref{t-subaru}).
The topological signal is clearly very strong in Fig.~\ref{f-quadruples},
both for low $\Delta t = 0.01$~{\hGyr} (top), in which case it is unlikely that 
stellar evolution would cause the earlier image of a galaxy to dim
too much to be seen, and for high $\Delta t = 1$~{\hGyr}, in which case
dimming could weaken the test. The northern part of each catalogue analysed consists
of the 44 known objects, there are 41 topologically implied objects within 
the southern field, and the simulated southern subset consists of 24 and 30 objects
for $\Delta t = 0.01$~{\hGyr} and 1~{\hGyr}), respectively.

\fquadrupleseuclidphotz

The uncertainty in the $T^3$ coordinates and fundamental length also
need to be taken into account. An error of $2^\circ$ in the vector (in
the flat comoving covering space) from the observer to an object at a
matched disc centre implies an error of about twice the size in the
observing angle of the implied image, because the diameter of the
shell containing objects at the distance of the matched disc centres
is twice the radius of the shell. Thus, as
Fig.~\ref{f-quadruples-T3-perturb} shows, a typical realisation of a
1~deg$^2$ southern survey is very unlikely to detect any topological
quadruples (satisfying the filter criteria). The total disappearance
of the topological signal in this figure ($s=0$ for all objects, since
there are no topological images lying within the southern field at all),
making the $N(s)$ distribution even weaker than that of the simply
connected data set, is an artefact of the method. We did not add any
``background'' simulated data around the northern galactic, observed
SDF field, so the implied southern objects constitute a ``survey''
that contains less objects (zero) than the simulated southern survey (16 objects in 
Fig.~\ref{f-quadruples-T3-perturb}).
Thus, let us consider fully simulated surveys.


\subsection{Planned or possible observations}
\label{s-results-planned} 


Given the uncertainties in the $T^3$ solution discussed 
in \SSS\ref{s-method-T3-WMAP} and listed in Table~\ref{t-T3-lb}, let us consider
a pair of northern and southern surveys centred on the matched disc centres
for the same $T^3$ axis, each of 64~deg$^2$. As explained above, for a given
known object, a two degree $T^3$ axis 
uncertainty yields an approximately four degree uncertainty 
for an observer placed halfway between the opposing images and searching for
a topological image at the expected position. However, since we simulate both
fields, other pairs of objects in the matched discs can be found, so $2^\circ$
should approximately correspond to a 68\% chance (for one angle offset from
a Gaussian distribution of width $2^\circ$) of finding matching pairs. 
Thus, 64~deg$^2$ should approximately correspond to a 2$\sigma$ 
(where $\sigma$ is one standard deviation) chance of finding topological
pairs.

\fconflevel

\fconflevelzfourzseven

The uncertainty in $\LTthr$ gives a 1$\sigma$ uncertainty of about
150~{\hMpc}, i.e. a 2$\sigma$ uncertainty of 300~{\hMpc}. For a survey
going from the matched disc centre to $\beta \approx 4^\circ$ away, 
Table~\ref{t-zmin} (for $\Omm=0.28$) gives a typical redshift of 
$z \approx 5.38$, i.e. $5.706$~{\hGpc} in radial comoving distance from
the observer. Inverting this, $5706 \pm 300 $~{\hMpc} gives
$ 4.60 \ltapprox z \ltapprox 6.33$. We extend this a little to 
be conservative, and increase the filter criterion 
$\epsilon$ to 2~{\hMpc}. A typical realisation is shown in the top
panel of Fig.~\ref{f-quadruples-euclid-photz}. Clearly, a pair of surveys that
are wide enough on the sky and in redshift coverage, finding 300 
objects above an idealised, $z$-independent completeness limit, has
a good chance of strongly rejecting or supporting the $T^3$ hypothesis.

Could photometric redshift estimates be sufficient? 
The lower panel of Fig.~\ref{f-quadruples-euclid-photz} shows 
the quadruple signal for 
a realisation with Gaussian peculiar velocity uncertainties
simulated with $\sigma(\betapec) = 0.005$, 
i.e. a redshift uncertainty of $\sigma(\Delta z) \approx 0.03$.
The signal is clearly much weaker, but (in this realisation) is still
distinguishable from the $N(s)$ curve of the simply connected simulation.

Figure~\ref{f-conf-level} presents confidence levels, i.e. $\probbetabest$,
as defined in Eq.~(\ref{e-defn-probbetabest}) in \SSS\ref{s-method-beta-alpha},
from ensembles of realisations.
Unsurprisingly, low $\sigma(\Delta \zphot)$ and high $N_{\mathrm{gal}}$ 
generally give the statistically most significant results.
The dependence on these two parameters is not fully monotonic. This is
understandable because for a fixed $\sigma(\Delta \zphot)$,
higher $N_{\mathrm{gal}}$ not only increases the numbers of topological
quadruples, it also increases the number density, so that the numbers
of non-topological quadruples also increase,
yielding some complexity in the dependence. There is some statistical
noise in the contours. 
The total computing time for Fig.~\ref{f-conf-level} on a single, recent 4-core processor
would be from a few weeks to a few months
(Appendix~\ref{a-bench}). The actual calculations were
performed using parallel computing resources on several
different machines.

Figure~\ref{f-conf-level} also shows that a typical photometric
redshift error of $\sigma(\Delta \zphot) \ltapprox 0.01$ would enable
rejection at $\probbetabest \ltapprox 0.05$ (i.e. a $2\sigma$
rejection according to intuition for a 
Gaussian distribution), provided that
$N_{\mathrm{gal}} \gtapprox 500$ in northern and southern surveys
each over $\omega \approx 100$~deg$^2$. This is only a moderate confidence level for
rejecting the hypothesis, but since $\probbetabest$ also represents
the expectation value of falsely accepting the hypothesis, it would be
sufficient to provide a strong motivation for studying candidate
topologically lensed galaxies further.  A plot roughly similar to the lower
panel of Fig.~\ref{f-quadruples-euclid-photz} would be obtained,
enabling the particular galaxies most likely to be topologically
lensed to be identified.  The particular realisation in the lower panel of
Fig.~\ref{f-quadruples-euclid-photz} shows about 10 galaxies that are
each members of $s > 1$ quadruples. This is a small enough number for
spectroscopic followup to be relatively easy to obtain. 

Comparison of the top and bottom panels in Fig.~\ref{f-conf-level} illustrates
the potential role of evolutionary effects. The lower panel limits
type II pairs to those with $\Delta t = 0.1$~{\hGyr} instead of
$\Delta t = 1$~{\hGyr} (upper panel). To attain moderate confidence,
i.e. $\probbetabest \ltapprox 0.05$, 
photometric redshift errors would need to be tightened 
to about $\sigma(\Delta \zphot) \ltapprox 0.005$, for slightly higher
numbers of galaxies.
For initial starburst durations not much shorter 
than a typical galaxy dynamical time of $\sim 1$~{\hGyr},
$\Delta t = 0.1$~{\hGyr} should give a high probability that a galaxy
is included in a survey at both its topological images.

Figure~\ref{f-conf-level-zfourzseven} shows that increasing
the survey area and redshift depth still further, to 
$\omega \approx 196$~deg$^2$ and $4 < z <7$, would allow statistically
similar results for weaker accuracy of the photometric redshifts
and lower numbers of galaxies. An accuracy of
$\sigma(\zphot) \ltapprox 0.02$ or
$\sigma(\zphot) \ltapprox 0.01$, 
for $\Delta t =1$~{\hGyr} (upper panel) or
$\Delta t = 0.1$~{\hGyr} (lower panel),
respectively, would give  $\probbetabest \ltapprox 0.05$. 

Figure~\ref{f-conf-level-zfourzseven} also shows that for
a sufficiently high number of galaxies, 
$N_{\mathrm{gal}} \gtapprox 900$, 
a photometric redshift accuracy of $\sigma(\Delta \zphot) \ltapprox 0.01$
would be sufficient for $\probbetabest \ltapprox 0.01$, i.e.
an expectation value of false rejection of about 1\%.
Most of the present interest in photometric redshift techniques is at
lower redshifts, but high statistical accuracy in $z \sim 6$ 
photometric redshifts would clearly be useful for deep redshift
topological lensing.

\section{Conclusion} \label{s-conclu}

Over the next few decades, wide-angle surveys in the $z \sim 6$
redshift range will inevitably be performed. The results above
indicate that with appropriate targetting and choices of observational
parameter limits, the speed with which the $T^3$ candidate for the
topology of the Universe can be rejected or detected can be optimised.
For more specific observational strategies for particular telescope/instrument
combinations, 
more detailed analyses could potentially \prerefereechanges{be} carried out by using 
detailed galaxy formation models, such as the hybrid
$N$-body/semi-analytic simulations
\citep{1993ASPC...51..298R,RPQR97,RNinin01} that have been extensively
developed to simulate detailed galaxy properties as a function of space and time 
\citep[e.g.,][and references thereof]{Hatton03}, 
including a specific focus on LAEs \citep[e.g.,][]{Garel12}.
More sophisticated combinations of existing observations, including
the VVDS~1400+05 field within 16$^\circ$ of the same southern axis 
and the VVDS~0226$-$04 field within 20$^\circ$ of the corresponding northern
axis (Table~\ref{t-deepfields}) along with simulated future observations,
should also potentially yield several alternative strategies for 
topological lensing detection or rejection.

The total number of $L_*$ galaxies at $4 \le z \le 7$, where $L_*$ is
the characteristic luminosity of a Schechter function
\citep{Schechter76}, is estimated to be $\phi_* \sim 3 \times
10^{-3}${\hMpccube} \citep[e.g. Table~5,][]{Bouwens11}.  This gives
about 3--8 million $L_*$ galaxies for the pairs of 100~deg$^2$ and
196~deg$^2$ fields simulated above, over $4.3 < z < 6.6$ and $4<z<7$,
respectively.  For the minimum $N_{\mathrm{gal}} \gtapprox 500$ or
$N_{\mathrm{gal}} \gtapprox 400$ needed for achieving
5\% expected confidence levels 
in the two field sizes, respectively 
(\SSS\ref{s-results-planned}),
surveys complete to $L \gtapprox 8.8 L_*$ or $L \gtapprox 9.9 L_*$,
respectively, would find the required numbers of galaxies. 
Because the computing time for the above calculations scales roughly as
$N_{\mathrm{gal}}^4$  
for sufficiently high $N_{\mathrm{gal}}$
(since the heaviest computation is checking the
list of possible quadruples; see Appendix~\ref{a-bench}), simulations 
for surveys complete to $L \gtapprox L_*$, i.e. 
for $N_{\mathrm{gal}} \gtapprox
10^6$, are clearly not practical without further filters added
early in the successive filter algorithm.

Nevertheless, the most difficult aspect of predicting 
\postrefereechanges{constraints} on the
model is the difference between {\em precise} cosmology and {\em accurate}
cosmology. The dark energy parameter $\Omega_\Lambda$ is suspected 
\postrefereechanges{by several cosmologists}
to
be an artefact of using the FLRW metric
(a homogeneous solution of the Einstein equation) 
rather than the physical average metric of the actually 
inhomogeneous Universe
\postrefereechanges{\citep[e.g.][]{BuchCarf03,Buchert08status,CBK10,WiegBuch10,Kolb11FOCUS,BoehmRasan13,Wiltshire12Hflow,BuchRZA2,ROB13}}.
A significant confirmation of the $T^3$ hypothesis would provide a constraint
on the inhomogeneous approach to cosmology, but this is unlikely to occur
using an FLRW interpretation of the observational data
if the time dependence of the FLRW metric parameters is too far from  
relativistically consistent formulae in the 
relevant redshift range \citep{Larena09template}. 
Some of the familiar FLRW relations may be algebraically valid
in fully relativistic models, with \prerefereechanges{an effective rather than a local} physical interpretation \citep{BuchRas12}, so detection
of topological lensing might still be possible under the assumption
of an FLRW metric.

\section*{Acknowledgments}

Thank you to Roland Bacon for contributing several key ideas to this
project, \postrefereechanges{and to the referee Andrew Jaffe for several
constructive comments}.
Some of this work was carried out within the framework of the European
Associated Laboratory ``Astrophysics Poland-France''.
\postrefereechangesbis{Part of this work consists of research conducted in the scope of the HECOLS International Associated Laboratory.} 
BFR thanks
the Centre de Recherche Astrophysique de Lyon for a warm welcome
and scientifically productive hospitality.
A part of this project has made use of 
Program Oblicze\'n Wielkich Wyzwa\'n nauki i techniki (POWIEW)
computational resources (grant 87) at the Pozna\'n 
Supercomputing and Networking Center (PCSS).
A part of this work was conducted within the ``Lyon Institute of
Origins'' under grant ANR-10-LABX-66.
Use was made of the Centre de Donn\'ees astronomiques de Strasbourg 
(\url{http://cdsads.u-strasbg.fr}),
the GNU {\sc plotutils} graphics package,
and
the GNU {\sc Octave} command-line, high-level numerical computation software 
(\url{http://www.gnu.org/software/octave}). 
\postrefereechangesbis{We acknowledge the use of the Legacy Archive for Microwave
     Background Data Analysis (LAMBDA), part of the High Energy
     Astrophysics Science Archive Center
     (HEASARC). HEASARC/LAMBDA is a service of the Astrophysics
     Science Division at the NASA Goddard Space Flight
     Center. This research has made use of the NASA/IPAC
     Extragalactic Database (NED) which is operated by the Jet
     Propulsion Laboratory, California Institute of Technology,
     under contract with the National Aeronautics and Space
     Administration. This research uses data from the VIMOS VLT
     Deep Survey, obtained from the VVDS database operated by
     Cesam, Laboratoire d'Astrophysique de Marseille, France.
     This research is based in part on data collected at Subaru
     Telescope, which is operated by the National Astronomical
     Observatory of Japan.}

%
%


\subm{ \clearpage }

%



\appendix
\section{Combination of cosmological and peculiar velocities in the FLRW model}
\label{a-velocity-addition}
As derived in \citet{Synge60} and presented in \citet{Narlikar94redsh} 
\citep[see also][and refs therein]{Rmovestationary10}, the expansion redshift 
can be derived by parallel-transporting the four-velocity of the world line of a distant
galaxy along a null geodesic (path of a photon) joining it to the observer. 
A fundamental observer (at rest with respect to the comoving spatial coordinate system) 
has zero peculiar velocity and a redshift of $\zcosm$.
The latter can be interpreted as a special-relativistic radial speed $\betacosm$ in natural units,
where
\begin{eqnarray}
  1 + \zcosm 
  &=& \sqrt{ \frac{1+\betacosm}{1-\betacosm} } \nonumber \\
  &=&  \frac{1}{\gammacosm(1-\betacosm)}   \nonumber \\
  &=&  \gammacosm(1+\betacosm)   \nonumber \\
  &=& \cosh \phicosm + \sinh \phicosm   \nonumber \\
  &=& \mathrm{e}^{\phicosm}, 
  \label{e-z-beta-phi-cosm}
\end{eqnarray}
where $\phicosm$ is the rapidity defined by $\betacosm =: \tanh \phicosm$, 
$\gammacosm := (1-\betacosm^2)^{-1/2},$ and
$\gammapec := (1-\betapec^2)^{-1/2}.$
For a radial peculiar velocity of $\betapec$ in natural units, the overall redshift $z$ 
can be calculated using
Minkowski spacetime addition of the rapidities $\phicosm$ and  $\phipec$,
where $\betapec =: \tanh \phipec$, i.e. the overall rapidity is
\begin{equation}
  \phi = \phicosm + \phipec,
\end{equation}
since addition of four-velocity vectors at the same spacetime location is meaningful.
Thus, similarly to Eq.~(\ref{e-z-beta-phi-cosm}), the overall redshift is given by
\begin{eqnarray}
  1 + z 
  &=& \mathrm{e}^\phi \nonumber \\
  &=& \mathrm{e}^{\phicosm + \phipec}  \nonumber \\
  &=& (1+\zcosm) (1+\zpec) ,
  \label{e-z-mult}
\end{eqnarray}
i.e. 
\begin{equation}
  z = \zcosm + \zpec + \zcosm \zpec .
  \label{e-z-mult-simpler}
\end{equation}
When $\max( \zcosm, |\zpec| ) \ll 1$, this reduces to $ z = \zcosm +
\zpec$ to first order in both redshifts.

For high-redshift astronomy, i.e. $\zcosm > 1$,
the third term in the right-hand side of 
Eq.~(\ref{e-z-mult-simpler}) is more significant than the second,
contrary to popular belief that sets the third term to zero.
Nevertheless, for $|\zpec| \ll 1$, which is realistic even for low-redshift
astronomy, Eq.~(\ref{e-z-mult-simpler}) 
can be approximated to first order in $\betapec$,
\begin{equation}
  z \approx \zcosm + \betapec + \zcosm \betapec .
  \label{e-z-mult-firstorder-pec}
\end{equation}
We ignore gravitational redshift here, since observations close to the
Schwarzschild radius are unlikely in the case of interest.
Parallel transport of the four-velocity \citep[see][]{Narlikar94redsh}
implies analogous relations to those above.

\fbench

\section{Simulation benchmarking} \label{a-bench}

Figure~\ref{f-bench} shows that for sufficiently high
$N_{\mathrm{gal}}$, the successive filter quadruple searches scale
roughly as the fourth power of the number of simulated galaxies. From
top to bottom as labelled, the power law best fits are proportional to
$N_{\mathrm{gal}}^{4.2},$ 
$N_{\mathrm{gal}}^{4.1},$ \prerefereechanges{and 
$N_{\mathrm{gal}}^{3.9},$}
respectively.  \prerefereechanges{Parallelisation} is via
\prerefereechanges{{\sc openmp}.}


\end{document}